\documentclass[aps,twocolumn]{revtex4}

\usepackage{graphicx}
\usepackage{color}
\usepackage{amsmath}
\usepackage{bm}

\begin{document}
\newcommand{\bea}{\begin{eqnarray}}
\newcommand{\eea}{\end{eqnarray}}
\newcommand{\be}{\begin{equation}}
\newcommand{\ee}{\end{equation}}
\newcommand{\bi}{\bibitem}
\newcommand{\D}{\mbox{$\Delta$}}
\newcommand{\mitD}{\mbox{${\mit \Delta}$}}
\renewcommand{\d}{\mbox{$\delta$}}
\renewcommand{\t}{\mbox{$\theta$}}
\renewcommand{\o}{\mbox{$\omega$}}
\renewcommand{\a}{\mbox{$\alpha$}}
\newcommand{\s}{\sigma}
\newcommand{\g}{\mbox{$\gamma$}}
\newcommand{\ep}{\mbox{$\epsilon$}}
\newcommand{\p}{\mbox{$\psi$}}
\newcommand{\bfk}{\mbox{${\bf k}$}}
\newcommand{\bfR}{\mbox{${{\bf R}}$}}
\newcommand{\bfr}{\mbox{${\bf r}$}}
\newcommand{\naR}{\mbox{$\nabla_{\bf R}$}}
\newcommand{\xipa}{\mbox{$\xi_{\parallel}$}}
\newcommand{\xipe}{\mbox{$\xi_{\perp}$}}
\newcommand{\xia}{\mbox{$\xi_{ani}$}}
\newcommand{\rr}{{\bf r}}
\newcommand{\kk}{{\bf k}}
\newcommand{\pp}{{\bf p}}
\newcommand{\qq}{{\bf q}}
\newcommand{\BSCCO}{{Bi$_2$Sr$_2$CaCu$_2$O$_8$ }}
\newcommand{\YBCO}{{YBa$_2$Cu$_3$O$_{7-\delta}$ }}
\newcommand{\nn}{\nonumber}
\newcommand{\G}{{\cal G}}
\newcommand{\bom}{{\bar\omega}}
\newcommand{\KK}{{\cal K} }
\newcommand{\EE}{{\cal E} }
\newcommand{\sgn}{{\rm sgn}}
\newcommand{\uG}{\underline{G}}
\newcommand{\ucG}{\underline{{\cal G}}}
\newcommand{\uGo}{\underline{G}^0}
\newcommand{\ug}{\underline{g}}
\newcommand{\uL}{\underline{L}}
\newcommand{\uM}{\underline{M}}
\newcommand{\uT}{\underline{T}}
\newcommand{\ut}{\underline{t}}
\newcommand{\uV}{\underline{V}}
\newcommand{\uVm}{\underline{V_1}}
\newcommand{\uU}{\underline{U}}
\newcommand{\uR}{\underline{R}}
\def\k{{\bf k}}
\def\r{{\bf r}}
\def\R{{\bf R}}
\def\p{{\bf p}}
\def\v{{\bf v}}
\def\D{{\cal D}}

\title{Electronic structure of $d$-wave superconducting quantum wires }

\author{A. M. Bobkov$^{1,2}$, L.-Y. Zhu$^{3}$, S.-W. Tsai$^{3, 4}$,
T. S. Nunner$^{3}$, Yu. S. Barash$^{1,2}$ and P. J. Hirschfeld$^{3}$\\~}

\affiliation{$^1$Lebedev Physical Institute, Leninsky Prospect 53,
Moscow 119991, Russia\\$^2$Institute of Solid State Physics,
Chernogolovka, Moscow reg., 142432 Russia \\$^3$Department of
Physics, University of Florida, Gainesville, FL
32611\\$^4$Department of Physics, Boston University, Boston, MA
02215}

\date{\today}

\begin{abstract}
We present analytical and numerical results for the electronic
spectra of wires of a d-wave superconductor on a square lattice.
The spectra of Andreev and other quasiparticle states, as well as
the spatial and particle-hole structures of their wave functions,
depend on interference effects caused by the presence of the
surfaces and are qualitatively different for half-filled wires
with even or odd number of chains. For half-filled  wires with an
odd number of chains $N$  at (110) orientation, spectra consist of
$N$ doubly degenerate branches. By contrast, for even $N$ wires,
these levels are split, and all quasiparticle states, even the
ones lying above the maximal gap, have the characteristic
properties of Andreev bound states. These Andreev states above the
gap can be interpreted as a consequence of
an infinite sequence of Andreev reflections experienced by
quasiparticles along their trajectories bounded by the surfaces of
the wire. Our  microscopic results for the local density of states
display atomic-scale Friedel oscillations due to the presence of
the surfaces, which should be observable by scanning tunneling
microscopy. For narrow wires the self-consistent treatment of the
order parameter is found to play a crucial role.  In particular,
we find that for small wire widths the finite geometry may drive
strong fluctuations or even stablilize exotic quasi-1D pair states
with spin triplet character.
\end{abstract}


\maketitle

\section{Introduction}
Since the discovery of high temperature superconductivity (HTS), the
origin of the pairing phenomenon in these materials has been the
subject of intense debate, and is still not clarified. Part of the
unusual nature of HTS which has hindered theoretical analysis is the
short coherence length, which allows short-wavelength fluctuations of
various types of local order to coexist with superconductivity. Most
probes of the nature of the superconducting state have been
restricted, until fairly recently, to measurements of bulk properties
and tunnelling through relatively large areas. Following the
pioneering work of Hess et al.\cite{Hess91}, it was realized that
scanning tunneling microscopy (STM) could provide an atomic-scale
picture of the superconducting state, particularly useful when
applied to inhomogeneous situations like the vortex lattice.
Measurements of this type were subsequently performed on
high-temperature superconductors \cite{Fischer95}. In the past few
years, scanning tunnelling microscopy on the surface of HTS have
compiled a novel and fascinating picture of the local electronic
structure of a few of these materials
\cite{yazdani,davisnative,davisZn,cren,davisinhom1,delozanne,howald,davisinhom2}.
In the Ba$_2$Sr$_2$CaCu$_2$O$_8$ system (BSCCO), one dramatic
implication of these experiments is that even relatively high
quality single crystals display inhomogeneous electronic structure
at the nanoscale \cite{davisinhom1,howald,delozanne,davisinhom2}.

In parallel to studies of the HTS materials, point contact
spectroscopy has been used to study the electronic structure of
ultrasmall  conventional superconducting
islands \cite{vonDelftreview}. Among many fascinating consequences
of the nanoscale geometry are number parity effects, in which the
qualitative electronic structure depends sensitively on whether
the number of electrons on the island are even or odd. More
recently,  superconducting wires of widths tens of
nanometers \cite{Tinkham} have also been fabricated.  Although
they have not yet been studied by STM or similar methods, this
should be technically feasible.

Isolated nanoscale grains and wires of HTS material have not been
fabricated to our knowledge. While this may prove technically
quite difficult to achieve due to the complexity of the crystal structure,
there seems to be no fundamental obstacle in the
long run.  This applies as well to other superconductors thought
to manifest unconventional superconducting order, where
effects of finite geometry should be easier to see since
coherence lengths tend to be larger.

It is our purpose in this paper to study how $d$-wave (e.g., HTS)
and other unconventional symmetry superconductors behave in finite
geometry at the atomic scale. Ziegler et al.\cite{zieglerwire}
began the study of this problem in the case of (100) $d$-wave
quantum wires, pointing out the dependence of the Fermi level
density of states on the parity of the wire width.   This is a
natural consequence of the discretization of the electronic energy
levels due to the finite wire width in the $d$-wave state.
While in the $s$-wave case all the interesting physics is tuned by
the level discretization, one expects a priori one  fundamental
difference in the $d$-wave case: for any geometry with surfaces
making an arbitrary angle with the crystal axes, pair-breaking
processes take place on a scale of the coherence length.
The most important consequence for the electronic structure should
be the formation of Andreev surface states.

However, little is known about how Andreev surface states behave
when the size of the superconductor becomes comparable to $\xi_0$.
The zero-energy states form on surfaces with orientations different
from the antinodal directions of a $d$-wave superconductor, due to
the sign change of the order parameter. In high-temperature superconductors,
such states manifest themselves as the zero-bias conductance peak
(ZBCP) in tunneling spectroscopy in the ab-plane \cite{geerk88,hu94,tan95,%
mats95,buch95,tan952,cov96,xu96,fog97,bbs97,cov97,alf97,ek97,ap98,alf981,%
alf982,sin98,wei98,apr99,deutsch99,Ting110,cov00,pairor02,greene02,wumou03,%
greene03}, the anomalous temperature behavior of the Josephson critical
current~\cite{tan96,bbr96,ilichev01,blamire03} and the upturn in the
temperature dependence of the magnetic penetration depth~\cite{walter98,%
bkk00,carr01} (see also review articles \cite{tan00,wendin01}).
The conventional description of Andreev surface states, as well as
the Andreev reflection itself, is based on the quasiclassical
approximation, a powerful tool in the theoretical study of various
properties of inhomogeneous superconducting systems.

The quasiclassical theory of superconductivity gives a so-called
coarse-grained description of the phenomena, averaged over
interatomic distances. This has been used, for example, to
calculate the local quasiparticle density of states (LDOS).
However, these coarse-grained averaged results are not adequate to
analyze atomic resolution measurements by STM and some other
contemporary experimental techniques (e.g. atomic force microscopy
\cite{giessibl}). To obtain this type of information, a fully
quantum mechanical atomic-scale approach going beyond the
quasiclassical approximation in describing inhomogeneous states of
superconductors is required. We address this problem in the
present paper using a tight-binding BCS-like model of $d$-wave
superconductor on a square lattice.

As a first step towards understanding the effect of constrained
geometry, we study the simplest case of $d$-wave superconducting
wires  consisting of $N$ parallel chains as the system size is
reduced. The  quasiparticle spectrum of such systems is described
both analytically, with an assumed spatially homogeneous order
parameter, and also numerically with a fully self-consistent
approach. In the limit $N\gg 1$ the usual surface Andreev states
in lattice models \cite{Ting110,tan00,pairor02}, as well as the
surface states known in continuous models \cite{bbs97}, can be
recovered at each surface of the wire. However, for sufficiently
narrow wires, when the transverse wire dimension is the order of
the superconducting coherence length, the Andreev states strongly
interfere and give rise to qualitatively new effects. We show
below that only those effects which occur for bands sufficiently
far from half-filling and relatively wide wires  can be described
with the quasiclassical theory of superconductivity. In addition,
we demonstrate how and under what  conditions one can recover
earlier quasiclassical results for Andreev states in $d$-wave
superconducting films \cite{nagai} from our microscopic approach.

The microscopic method adopted in this paper to constrain the
 geometry involves introducing lines of impurities of
potential strength taken to infinity to bound the wire.
 We  show that earlier microscopic results on (110) surfaces
 \cite{bbs97,Ting110,pairor02,Walker,tan00} and (100) wires
\cite{zieglerwire} can be reproduced by this technique, and then
 extend it to calculate new results on  wires with other
 orientations.  We find that the results for electronic spectra
 are very sensitive to the number parity of the wire width, and
 that true zero-energy Andreev states can only exist in wires with
 odd numbers of chains.  In even wires, the Andreev states are
 split,
 pushed away from the Fermi energy, and can have either surface or
 standing wave character.   Finally, we show that for smaller
 wires self-consistency effects become important and can even,
 within mean field theory, lead to condensation of a fully gapped
 spin triplet state instead of $d$-wave order.

Our results may also have some qualitative relevance for the
related problem of $d$-wave superconducting grains, where the
geometry is constrained in 2 dimensions.  In addition to
artificially fabricated islands, some authors have proposed that
the BSCCO-2212 samples which display nanoscale inhomogeneity
should be thought of as a collection of weakly coupled $d$-wave
grains of roughly the size of the superconducting coherence length
$\xi_0$, or $d$-wave grains coupled to grains of another
electronic phase \cite{davisinhom1,howald,davisinhom2,Joglekar}.
In fact, the structure of a general, possibly irregular small
grain of $d$-wave superconductor has not been studied to our
knowledge. Understanding how the local density of states (LDOS) of
these wires depend on the wire width and the orientation, as well
as on the deviation from half-filling, could provide important
intuition for the question of the electronic structure of the
small irregular grains possibly present in BSCCO samples.

The outline of the paper is as follows. In Sec.\ref{sec:Tmatrix},
we introduce the formalism for the problem.  In
Sec.\ref{sec:surface}, we discuss three special semi-infinite
surface orientations: $(110)$, $(210)$ and $(100)$. In Sec.\ref
{sec:swires}, we study the $(110)$ superconducting wires with even
and odd width and make an comparison with the discrete states in
normal metal wires (Sec.\ref{sec:nwires}) to try to identify the
nature of the true Andreev states. In Sec.\ref{sec:finitemu}, we
study the effects of deviations from the half-filling. In
Sec.\ref{sec:self} the results of fully self-consistent
calculations are presented. In particular, we allow the order
parameter to vary spatially and comment on the differences in our
results. In the case of narrow wires the self-consistent study
shows the appearance of some peculiar types of superconducting
pairing. Finally, in Sec.\ref{sec:conclusion}, we present our
conclusions.

\section{Model description and formalism}
\label{sec:Tmatrix}
The Hamiltonian for a pure singlet superconductor
can be written as:
\begin{eqnarray}
\cal{H} &=& - t \sum_{\langle i,j\rangle, \sigma}
c^\dagger_{i\sigma} c_{j\sigma} - \sum_{i,\sigma} [\mu - U_i]
c^\dagger_{i\sigma} c_{i\sigma} \nonumber \\ &+& \sum_{\langle
i,j\rangle} \{ \Delta_{ij} c^\dagger_{i\uparrow}
c^\dagger_{j\downarrow} + h.c. \}, \label{ham}
\end{eqnarray}
where we have chosen a  nearest neighbor tight-binding band for
simplicity; $\mu$ is the chemical potential.
A superconducting pairing is defined for nearest neighbors
$\Delta_{ij} = -V \langle c_{j\downarrow}c_{i\uparrow}
\rangle$ on the bond $\{i,j\}$. The parameter $t$ is of order 150
meV for high $T_c$ materials, and we consider both the
particle-hole symmetric model $\mu=0$ and the more realistic case
of finite $\mu$. In non-self-consistent calculations the OP has
the familiar $k$-space form $\Delta_\k = \Delta_0 [ \cos(k_aa) -
\cos(k_ba) ]$, where $ \Delta_0 = \frac{1}{2}\sum_\pm (
\Delta_{i\,i\pm r_a} - \Delta_{i\,i\pm r_b} ) $ is independent of
$i$, and  is taken to be $0.2t$. The lattice constant is denoted by
$a$. The maximum gap is $\Delta_{max}=2\Delta_0$. We also present
self-consistent calculations, in which case $V$ is chosen to yield
this same value of $\Delta_0$ far from wire edges.

It is possible to constrain the geometry underlying Eq.
(\ref{ham}) in several different ways. We present results here for
a method discussed, for example, in Refs.\
\onlinecite{MorrDemler,Ting110} in which the on-site potentials
$U_i$ are chosen to lie on the boundary and their value is taken
to infinity to cut off electron transport through the boundary.
This technique has the virtue that the strength of the barrier can
in principle be lowered to allow different degrees of transparency
and the study of tunneling phenomena. In this work we restrict
ourselves to impurity configurations and strengths for which the
constrained system is completely isolated from its environment.
This is equivalent to the assumption of open boundaries, when no
hopping and no pairing take place outside the region. No disorder
is introduced in the system and we consider $U_i$ as ``impurity''
potentials only so as to form the surfaces of the superconducting
region.

The Hamiltonian of the surface term is taken to be \be U
=U_0\sum_{\ell} c_\ell^\dagger c_\ell, \ee where the set of sites
$\ell$ is determined exclusively by the boundaries of the desired
system (see below). The full Fourier space Green's function for
the system in the presence of these impurities is quite generally:
\begin{eqnarray}
\check{G}(\k,\k^{\prime},\omega) &=& \check{G}^{(0)}(\k,\omega)
\delta_{\k,\k^{\prime}} +\nonumber\\
&+&\check{G}^{(0)}(\k,\omega)
\check{T}(\k,\k^{\prime},\omega) \check{G}^{(0)}(\k^{\prime},\omega)
\enspace ,
\label{greens}
\end{eqnarray}
where the $T$-matrix can be found from the following equations:
\begin{eqnarray}
\check{T}(\k,\k^{\prime},\omega)&=&\check{U}(\k,\k^{\prime})+\nonumber\\
&+&\sum_{\k^{''}} \check{U}(\k,\k^{''})
\check{G}^{(0)}(\k^{''},\omega)
\check{T}(\k^{''},\k^{\prime},\omega) . \label{tmat}
\end{eqnarray}
Here $\check{G}$ and $\check{T}$ take $4\times4$ matrix form in
the four-dimensional product space of particle-hole and spin
variables. If we choose nonmagnetic on-site potentials and
consider singlet superconductors, the problem reduces to $2\times
2$ matrices in Nambu space. The Nambu retarded propagator for the
pure $d$-wave superconductor is

\begin{equation}
\hat{G}^{(0)}(\k,\omega)={\omega \hat{\tau}_0 +\xi_\k \hat{\tau}_3
+ \Delta_\k\hat{\tau}_1 \over (\omega+i0)^2 -
\xi_\k^2-\Delta_\k^2} \enspace ,
\end{equation}
where the $\tau_\alpha$ are the Pauli matrices and
$\xi_{\bf k}=-2t[\cos(k_aa)+\cos(k_ba)]-\mu$.
Calculating the  $T$-matrix allows us to  obtain the
eigenenergies of the system from the condition
${\rm det }\hat{T}^{-1}(\omega)=0$.

The local spin-resolved quasiparticle density of states is given as
\begin{eqnarray}
    \rho_{\uparrow(\downarrow)}({\bf r},\omega) &=& -\pi^{-1}\mbox{Im }
    G_{11,\uparrow\uparrow(\downarrow\downarrow)}({\bfr},\r,
\omega) \enspace .
\end{eqnarray}
After integration of the LDOS over energy, we should
obtain the total number of quasiparticle states per one site.
Since each site on the lattice possesses 2 states with opposite
spins, the spin-resolved LDOS normalization is

\begin{equation}
\int_{-\infty}^\infty d\omega \rho_{\uparrow (\downarrow)}(\r,\omega) = 1.
\end{equation}

\section{surface case }

\label{sec:surface}

\begin{figure}[tbh]
\begin{center}
\leavevmode
\includegraphics[width=.6\columnwidth]{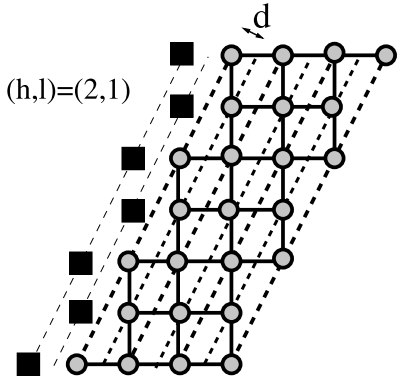}
\caption{(210) surface or wire. Two lines of impurities (closed
squares) are needed to isolate the lattice sites (closed circles)
with  nearest neighbor hopping.} \label{fig:surf210}
\end{center}
\end{figure}

Upon a conventional reflection on the (110) surface of a $d$-wave
superconductor, the order parameter always changes sign as the
direction of the quasiparticle momentum $\k$ is varied. This leads
to  Andreev reflection and, eventually, to the formation of the
dispersionless zero-energy surface Andreev bound states. For other
surface orientations the sign change does not take place for all
incoming momentum directions.
 It is important to notice that the number of consecutive
impurity lines that are needed in order to cut the system depends
on the orientation of the surface one wants to consider. For (100)
and (110) surfaces and nearest neighbor hopping, one line of
impurities is sufficient to cut communication between the two
sides. For a (210) surface, the nearest neighbor hopping and
pairing terms can still connect this particle with another across
a single impurity line, so for a simple tight-binding band,  two
consecutive lines are needed to close the system (Fig.
\ref{fig:surf210}). Alternatively, if one includes a next-nearest
neighbor hopping $t'$, even a (110) surface is not closed by a
single line of infinite impurities; the system considered in Ref.
\cite{MorrDemler} is therefore not a closed (impenetrable)
surface. For a ($hl0$) surface in general, max({$h,l$}) lines are
needed for a model that includes nearest-neighbor terms only, and
$h+l$ lines are needed for a model that includes
next-nearest-neighbor terms. Clearly, the technique becomes
cumbersome for arbitrary angles, but one can nevertheless learn a
good deal by considering special cases.

 In the presence of a surface of arbitrary orientation, the
simplest way of applying Bloch's theorem to this discrete system
is by using a surface-adapted Brillouin zone\cite{pairor02,Walker}.
We define new coordinates ($\hat{x}$, $\hat{y}$), rotated with
respect to the crystal axes ($\hat{a}$, $\hat{b}$), where
$\hat{x}$ is the direction normal to the surface and $\hat{y}$ is
the direction along the surface. The system is periodic
along the $y$-direction and the crystal momentum component $k_y$
of a quasiparticle is conserved. Instead of the usual
square Brillouin zone $k_a = [-\pi, \pi]$, $k_b = [-\pi, \pi]$
(for unit lattice constant $a=1$) we now use the surface-adapted
Brillouin zone given by $k_x =[-\pi/d, \pi/d]$ and
$k_y = [-\pi d, \pi d]$. Here $d =1/\sqrt{h^2+l^2}$ is the distance
between the nearest chains (layers) aligned along the surfaces.
The momenta in the two coordinate systems are simply related
through rotation of an angle $\theta = \tan^{-1} h/l$.

Now we turn to the solution of the equation for the $T$-matrix
(Eq. \ref{tmat}) for the case of one line of impurities. We start
with the ansatz
\begin{eqnarray}
\hat{T} = T_0 \hat{\tau}_0 + T_1 \hat{\tau}_1 + T_3 \hat{\tau}_3
\end{eqnarray}
and find, for arbitrary strength of impurity potential,
\bea T_0(k_y, \omega) &=& {{G}_0^{(0)}(0, k_y, \omega) \over
D_1} \ , \\
T_1(k_y, \omega) &=& {-{G}_1^{(0)}(0, k_y, \omega) \over D_1} \ , \\
T_3(k_y, \omega) &=& {c-{G}_3^{(0)}(0, k_y, \omega) \over D_1}
\eea
where $D_1(k_y, \omega) = [c-{G}_3^{(0)}(0, k_y, \omega)]^2 -
{G}_0^{(0)}(0, k_y, \omega)^2 + {G}_1^{(0)}(0, k_y, \omega)^2$, $c
= 1/V_0$  and ${G}_i^{(0)}(x, k_y, \omega)$ is the Fourier
transform with respect to $k_x$ of the $i$-th Nambu component of
the bare Green's function $G_i^{(0)}(\k, \omega)$
\bea
{G}_i^{(0)}(n, k_y, \omega) =\frac{d}{2\pi}
\!\!\int\limits_{k_x=-\pi/d}^{\pi/d}\!\! G^{(0)}_i (k_x, k_y, \omega) e^{i
k_x n d} dk_x\ .
\label{gtilde}
\eea
The site index $n$
corresponds to $x$-coordinate $x= nd$. For the case of infinitely
strong impurity potential considered here, $c = 0$. In this case
the expression for ${\hat T}$ can be written in the compact form \bea
\hat{T}(k_y, \omega) = - \left[\hat{{G}}^{(0)}(0, k_y, \omega)
\right]^{-1} \label{t} \eea and the poles of the $T$-matrix
correspond to zeros of the determinant of the Green's function
$\hat{{G}}^{(0)}(0, k_y, \omega)$.

The Fourier transform with respect to $k_x$ of Eq. \ref{greens} is
\bea &&\hat{{G}}(n, n^{\prime}, k_y, \omega) = \hat{{G}}^{(0)} (n
- n^{\prime}, k_y, \omega) -
\\ \nonumber
&&\ \ \ \ \ \ \  \ \ \hat{{G}}^{(0)}(n, k_y, \omega)
\left[\hat{{G}}^{(0)}(0, k_y, \omega)\right]^{-1}
\hat{{G}}^{(0)}(-n^{\prime}, k_y, \omega) \label{gn} \eea Due to
periodicity of the system along the $y$-direction, calculation
of the LDOS at site $n$ will simply involve a sum of ${G}(n,n,
k_y, \omega)$ over all values of $k_y$ within the surface-adapted
Brillouin zone and over two spin directions:
\begin{equation}
\rho(\r,\omega)=\rho(n,\omega)= -\frac{2}{\pi}{\rm Im} \int_{-\pi
d}^{\pi d}\frac{dk_y}{2\pi d} {G}_{11}(n,n,k_y,\omega)
\label{11}
\end{equation}

\subsection{(100) surface}

\begin{figure}[tbh]
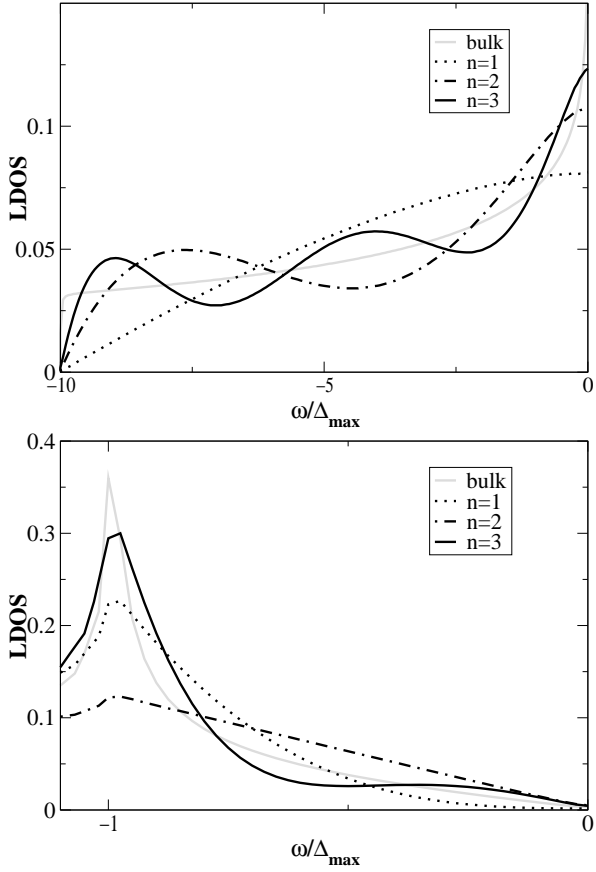

\begin{center}
\leavevmode
\includegraphics[clip=true,width=.9\columnwidth]{surface100_N.eps}
\includegraphics[clip=true,width=0.9\columnwidth]{surface100_S.eps}
\caption{ Local density of states $\rho(x,\omega)$ for a (100)
surface. Normal state (upper panel) and superconducting state (lower panel),
$\rho(x,\omega)$ vs. $\omega/\Delta_{max}$ for various
distances $n=x/d$ from surface; $\Delta_{max}=0.4t$, $\mu=0$.}
\label{fig:surf100}
\end{center}
\end{figure}

Since (100) surfaces are not pairbreaking in $d$-wave
superconductors, we do not {\it a priori} expect to see
interesting physics arising from Andreev states. On the other
hand, the mere presence of a surface can induce surface Tamm bands
\cite{Tamm}, decaying in the bulk on the atomic scale and experiencing
the Friedel-like oscillations of the LDOS. In our model Tamm states
have nothing to do with the superconductivity, although the
possibility for pairing of electrons occupying these surface
states is not excluded \cite{ginzburg}. While it is not our intent
to study these in detail, we present some results to show that
such states can be seen by the STM even in situations when surface
Andreev states are absent.  Fig.\ref{fig:surf100}  shows on upper
and lower panels the  LDOS for various distances from the surface
in the normal metal and the superconducting
states, respectively.

\subsection{(110) surface}

Results for a (110) surface are expected to be qualitatively
different from the ones obtained for a (100) surface. The bulk
order parameter in the coordinate system of the crystal axes $\hat
a$ and $\hat b$ is $\Delta_\k =\Delta_0(\cos k_x a -\cos k_y a)$,
but in the coordinates of the surface $\hat x$ (perpendicular to
surface) and $\hat y$ (parallel to surface) becomes $\Delta_\k = 2
\Delta_0 \sin k_x d \sin k_y d$, with $d=a/\sqrt{2}$.
 In this
case, an incident particle with any non-zero $k_y$ experiences a
sign change in the order parameter as it reflects from the
surface.

\begin{figure}[tbh]
\begin{center}
\leavevmode
\includegraphics[clip=true,width=.9\columnwidth]{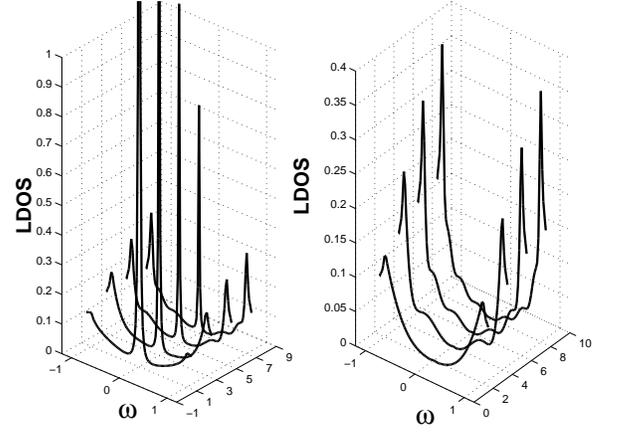}
\caption{Local density of states $\rho(x,\omega)$ in
superconducting state versus $\omega/\Delta_{max}$ for the case of
one (110) surface, $\Delta_0=0.2t$, $\mu=0$. Left: chains an {\it
odd} distance $n=x/d$ from surface at $x=0$. Right: chains an {\it
even} distance from $x=0$. } \label{fig:surf110}
\end{center}
\end{figure}

Calculation of the bare Green's function $\hat{{G}}^{(0)}(n, k_y,
\omega)$ from Eq. \ref{gtilde} gives at $\mu=0$, $G_{1,3}(n, k_y,
\omega)=0$ for even $n$ and $G_{0}(n, k_y, \omega) = 0$ for odd
$n$. Explicitly,
\begin{eqnarray}
\hat{{G}}^{(0)}(n=2m,k_y,\omega)=\qquad\qquad\qquad\qquad\qquad\qquad
&\nonumber
\\
\qquad\qquad -\frac{i|\omega|\exp\left\{-i|n|z\
\mathrm{sgn}\left[\omega(q^2-\Delta^2)\right]\right\}}
{\sqrt{(\omega^2-\Delta^2)(q^2-\omega^2)}}\hat{\tau_0}
\label{geven}
\end{eqnarray}
and \bea \hat{{G}}^{(0)}(n=2m+1,k_y,\omega)=\qquad \qquad
\qquad \qquad \qquad \qquad\nonumber \\
=\frac{i\Delta \exp\left[-i|n|z\
\mathrm{sgn}\left(\omega(q^2-\Delta^2)\right)\right]\mathrm{sgn}
\left(n(q^2-\Delta^2)\right)}
{\sqrt{(q^2-\Delta^2)(\omega^2-\Delta^2)}} \hat{\tau}_1\nn\\
+\frac{iq \exp\left[-i|n|z\
\mathrm{sgn}\left(\omega(q^2-
\Delta^2)\right)\right]\mathrm{sgn}(\omega)}{\sqrt{(q^2-\Delta^2)
(q^2-\omega^2)}} \hat{\tau}_3 \quad , \qquad \label{godd} \eea

where

\bea
z(k_y) &=& \tan^{-1}\sqrt{{q^2-\omega^2 \over \omega^2- \Delta^2 }}
\label{zomega} \enspace ,\\
\Delta(k_y) &=& 2 \Delta_0 \sin(k_y d) \label{dky} \enspace ,\\
q(k_y) &=& 4 t \cos(k_yd) \ , \label{qky} \eea and $\Delta(k_y)$
and $q(k_y)$ are the maximum gap $\Delta_\k$ and single-particle
spectrum $\xi_\k$ for fixed $k_y$ in 110 geometry. Note that here
the square root function takes positive values for positive
arguments, i.e. under the conditions
$|\Delta(k_y)|<|\omega|<|q(k_y)|$ or
$|q(k_y)|<|\omega|<|\Delta(k_y)|$. For $A^2-\omega^2<0$,\,the
branch $\sqrt{A^2-\omega^2}\to -i\ \mathrm{sgn}\omega\
\sqrt{\omega^2-A^2}$, for either $A=\Delta, q$.  In Fig.
\ref{fig:surf110} we show the LDOS spectra on the different
layers. Each layer is defined as an array of sites parallel to the
surface. Its index indicates its position; layer $n$ corresponds
to sites at a distance $nd$ away from the surface.

\begin{figure}[tbh]
\begin{center}
\leavevmode
\includegraphics[clip=true,width=.8\columnwidth]{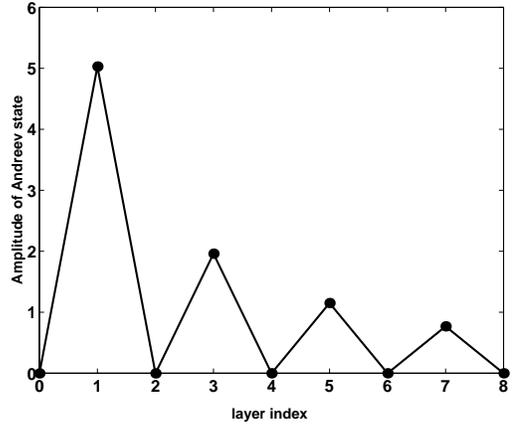}
\caption{Amplitude of Andreev state vs. distance $n=x/d$ from 110
surface for $\Delta_0=0.2t$, $\mu=0$ .} \label{fig:Andreev}
\end{center}
\end{figure}

The Andreev bound states are manifested as zero-energy peaks in
the LDOS. For $\mu=0$, such  states are found in all odd layers,
and are absent on even ones, as seen in Figure \ref{fig:Andreev}.
This even-odd effect can be easily understood from the form of the
$T$-matrix and Green's functions Eqs. (\ref{t}-\ref{godd}).
For even $n$, the bare Green's function ${G}^{(0)}(n,k_y,\omega)$
is proportional to $\omega$ for low-frequencies. Then the
$T$-matrix $T \sim 1/\omega$ for small $\omega$.  Even though the
$T$-matrix has a pole at $\omega=0$, the product of two Green's
functions in the analog of Eq. \ref{greens}  decreases faster,
resulting in zero LDOS at zero frequency. As seen from the right
panel of Fig. \ref{fig:surf110}, the low-energy density of states
on even layers near the $(110)$ surface is substantially less than
in the bulk, where it varies linearly with sufficiently low
energy. Furthermore, the amplitude of the gap features in the LDOS
is noticeably suppressed with decreasing distance from the
surface. These features of the LDOS can be understood based on a
simple relation between the Green's function ${\hat G}(n'=n, k_y,
\omega)$ of the half-space and the bare Green's function ${\hat
G}^{(0)}(n-n=0, k_y, \omega)$, which takes place for even $n=2m$ :
\begin{eqnarray}
&&\hat{{G}}(n'=n, k_y, \omega)=\Bigl\{1-\qquad \qquad \qquad
\qquad \qquad
\qquad \nonumber \\
&&-\exp\left[-2i|n|z\
\mathrm{sgn}\left(\omega(q^2-\Delta^2)\right)\right]\Bigr\}
{\hat G}^{(0)}(0, k_y, \omega) . \label{g2m}
\end{eqnarray}
The factor in the braces in Eq.(\ref{g2m}) controls the deviation
from the bulk behavior and diminishes with decreasing even $|n|$.
At low energies only narrow regions of $k_y$ near the center and
the edge of the Brillouin zone contribute to the LDOS. As a
result, for even layers the density of states goes as $|\omega|^3$
at low energies, with the main contributions arising from $k_y$
near the edge of the Brillouin zone.

For odd layers, the $\tau_1$ and $\tau_3$ components of
$\hat{{G}}^{(0)}(n=2m+1,k_y,\omega)$ are the ones that are
non-zero, and they approach a constant value as $\omega$ goes to
zero. So the pole in the $T$-matrix generates the peak in the
LDOS, associated with the zero-energy Andreev surface states. For
Andreev states, $z$ is an imaginary quantity. The size of the peak
decreases as the distance to the surface increases due to the
$e^{-|n {\mathrm Im} z|}$ factor in Eq.(\ref{godd}). For large
$|n|$, small $k_y$ dominates the integration over $k_y$ in the
LDOS and we obtain the following zero-energy asymptotic behavior
of the LDOS:
$\rho(\omega)=\dfrac{2t}{\pi\Delta_0n^2}\delta(\omega)$. The size
of the zero-energy peak in the LDOS $\propto n^{-2}$.

\subsection{$(210)$ surface}
It is useful to study a case intermediate between the standard
$(100)$ and $(110)$ surfaces to see what qualitatively new features
arise. From the usual quasiclassical viewpoint, the weight of the
zero-energy Andreev $(210)$-surface states should be finite, but
smaller than for the $(110)$ surface because the phase space for
which the reflecting quasiparticle experiences a sign change of the
order parameter is reduced. The tight-binding model leads
to a more complicated dependence of the weight of the zero-energy
states on the surface orientations relative to the crystal axes.
Thus, for the $(210)$ surface the model shows at half-filling
no zero-energy Andreev states at all. We associate this discrepancy,
in particular, with the difference between reflection channels
incorporated in the two approaches.

Standard quasiclassical considerations imply that parallel to a
smooth surface the momentum component $k_y$ is conserved in a reflection
event, and only conventional specular reflection takes place. A
tight-binding model shows that, generally speaking, this is not the
case, since the {\it crystal momentum} component $k_y$ can also change in a
reflection process by a reciprocal crystal vector along the surface.
Due to a difference between reciprocal crystal vectors at the surface
and in the bulk, the momentum acquired by a quasiparticle in a
reflection event can be physically distinguished in the bulk from
that of specularly reflecting quasiparticle. Hence, specific crystal
periodicity along a particularly oriented surface can result in
additional channels for quasiparticle reflection, if there is a reflected
state $k_{x}(k_y),\, k_y$ on the Fermi surface corresponding to the
$k_y$ surface Umklapp process \cite{gantmakher}.

In this case the Fermi surface, considered as a part of the surface
adapted Brillouin zone, should exhibit multiple values of outgoing
$k_{F,x}$ for a fixed value of $k_{F,y}$. This situation is realized
for the comparatively complicated multisheet structure of the Fermi
surface in the surface-adapted Brillouin zone for the (210) surface,
as obtained in Fig.9 of Ref.\onlinecite{pairor02}. A strong dependence
of the particular shape of the surface-adapted Brillouin zone, as well
as the Fermi surface in the zone, on surface orientation is an important
characteristic feature of the tight-binding models \cite{Walker,pairor02}.
Within the quasiclassical theory, the shape of the Fermi surface is usually
considered as independent of surface orientations relative to the
crystal axes. We note, that the zero-energy surface states, as a rule,
do not exist within the quasiclassical approach,
if multiple channels for reflection of quasiparticles from an impenetrable
surface are assumed. In this subsection, we now study the LDOS for
quasiparticle spectra obtained with the tight-binding model at half-filling
for (210) surface.

Technically, we now need  to solve for the Green's function in the
presence of two impurity lines (Fig. \ref{fig:surf210}).
In the general case of two parallel impurity lines, we cut the
crystal at an orientation given by ($hl0$) and introduce an
impurity potential
\bea
U(\r) = U_0 \sum_j \delta(\r- \R_j) \enspace ,
\eea
where $\R_j$ are the points on the two impurity lines, at the
location of the boundaries. The first boundary is defined to be
located at $x=0$ with impurity sites $\R_j = j d^{-1} \hat{y}$
and the other is parallel and located at $x=(N+1)d$ with
$\R_{j} = (N+1)d \hat{x} + (c+jd^{-1})\hat{y}$, giving a total
of $N$ free chains. Here $c$ is a shift along $y$-axis of sites
on the $(N+1)$-th chain relative to the sites on the $0$-th chain.
For the special case of the $(210)$ surface, we need two
adjacent lines at $x=0$ and $x=-d$, so formally this corresponds
to the case of $N=-2$.

The equation for the $T$-matrix (Eq. \ref{tmat}) for this impurity
potential can be solved by choosing the ansatz
\bea
\hat T(k_x,k_x^{\prime}, k_y) &=& \hat t_0 + \hat t_1 e^{i (N+1)d k_x} +
\hat t_2 e^{-i(N+1)d k_x^{\prime}} \nonumber \\
&&+ \hat t_3 e^{i(N+1)d (k_x-k_x^{\prime})} \enspace .
\eea
In the limit of infinitely strong impurity potential this gives
\bea
\hat t_0 &=& {-\hat{G}^{(0)}(0)^{-1} \over 1 - \hat{ G}^{(0)}(0)^{-1} \hat{
G}^{(0)}(-N-1)\hat{ G}^{(0)}(0)^{-1} \hat{ G}^{(0)}(N+1)} \nonumber \\
\hat t_1 &=& {\hat{ G}^{(0)}(0)^{-1} \hat{ G}^{(0)}(N+1) \hat{
G}^{(0)}(0)^{-1} \over 1 - \hat{ G}^{(0)}(0)^{-1} \hat{
G}^{(0)}(-N-1)\hat{ G}^{(0)}(0)^{-1} \hat{ G}^{(0)}(N+1)}\nonumber \\
\hat t_3 &=& - {\hat{ G}^{(0)}(0)^{-1} \over  1 - \hat{
G}^{(0)}(0)^{-1} \hat{ G}^{(0)}(N+1) \hat{ G}^{(0)}(0)^{-1} \hat{
G}^{(0)}(-N-1)} \nonumber \\
\hat t_2 &=& {\hat{ G}^{(0)}(0)^{-1} \hat{ G}^{(0)}(-N-1) \hat{
G}^{(0)}(0)^{-1} \over 1 - \hat{ G}^{(0)}(0)^{-1} \hat{
G}^{(0)}(-N-1) \hat{ G}^{(0)}(0)^{-1} \hat{ G}^{(0)}(N+1)}
\nonumber
\eea
The Green's function in this case can then be
written as:
$$ \hat{ G}\!(n, n^{\prime}) = \hat{G}^{(0)}\!(n-n^{\prime}) -
\left(\hat{ G}^{(0)}\!(n) \ \ \ \hat{
G}^{(0)}\!(n-N-1) \right)\times$$
\bea
\!\!\left(\!\begin{array}{cc}\! \hat{ G}^{(0)}\!(0) & \hat{ G}^{(0)}\!(
\!- N\!-\!1)\\
\hat{ G}^{(0)}\!( N\!+\!1) & \hat{ G}^{(0)}\!(0)
\end{array} \!\right)^{\!-1}\!
\!\left(\!\begin{array}{c} \hat{ G}^{(0)}\!(-n^{\prime})\\
\hat{ G}^{(0)}\!(\! N\!+\!1\!-\! n^{\prime})
\end{array}\!\right)\enspace ,
\label{greenwire}
\eea
where we have not written the
dependence on $k_y$ and $\omega$ explicitly.

\begin{figure}[tbh!]
\begin{center}
\leavevmode
\includegraphics[width=.9\columnwidth]{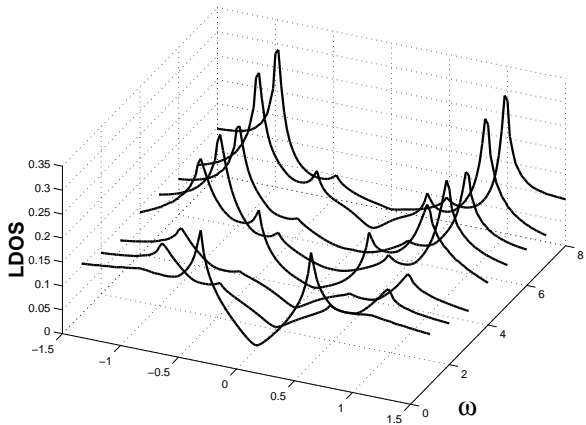}
\caption{Local density of states $\rho(x,\omega)$ versus
$\omega/\Delta_{max}$ for a closed(210) non-transparent surface at
half-filling model and $\Delta_0=0.2t$, $\mu=0$. Each curve
corresponds to a chain located at a distance $x/d$ to the
surface.} \label{fig:LDOS210}
\end{center}
\end{figure}

The equation for the bound state on (210) surface is
\bea
{\rm det} \left(\begin{array}{cc} \hat{{G}}^{(0)}(0) &
\hat{{G}}^{(0)}(1)\\
\hat{{G}}^{(0)}(-1) & \hat{{G}}^{(0)}(0) \end{array} \right) = 0
\enspace .
\label{det210}
\eea

Fig \ref{fig:LDOS210} displays the LDOS spectra calculated on
several layers for (210) surface.  There is no zero-energy peak
observed on any chain at all, for the reasons explained in the
beginning of this subsection. The peaks in the LDOS, seen close to
the (210) surface at energies $\pm 0.5\Delta_{max}$, originate
from the gap features taken for the momentum along the surface
normal: $\Delta(k_{F,x}, k_{F,y}=0)=0.5\Delta$. The peak at
$\omega=\Delta(k_{F,x},0)$ arises for the homogeneous model of
the order parameter, while for the self-consistent spatially
dependent order parameter it lies slightly below $\Delta(k_{F,x},0)$.
It is associated with the
surface Andreev states and decays in the depth of the
superconductor. These peaks have been theoretically found first in
Ref.\ \onlinecite{bbs97} with a continuous model and then also for
the conductance with a lattice model \cite{pairor02}. We notice
that the conductance spectrum shown in Fig.14 of
Ref.\onlinecite{pairor02}, is in agreement with the LDOS on the
first chain (at $x=d$) in Fig.\ref{fig:LDOS210}. The variations of
the LDOS from chain to chain, which accompany a large-scale
behavior, are the Friedel-like oscillations.

\section{Half-filled Wires}

We now consider wires where the second line of impurities confines
the system to a finite width, i.e. we restrict ourselves to the
cases with surface normal along the $(100)$ and $(110)$
directions. Semiclassically, a quasiparticle will go through
multiple scatterings, bouncing back and forth between the two
walls. Hence we expect that the interplay between Andreev
reflection, taking place due to sign change of the order
parameter, and the energy discretization, due to finite wire
width, to yield novel features. We again model the surfaces by
introducing impurities on the appropriate sites to completely
isolate the wires. The corresponding Green's function is given by
Eqs.(\ref{greenwire}). The bound state energies are determined by
the equation \bea {\rm det} \left(\begin{array}{ll}
\hat{{G}}^{(0)}(0) &
\hat{{G}}^{(0)}(-N-1)\\
\hat{{G}}^{(0)}(N+1) & \hat{{G}}^{(0)}(0) \end{array} \right) = 0
\enspace .
\label{det}
\eea

\subsection{(100) wires}

\begin{figure}[tbh]
\begin{center}
\leavevmode
\includegraphics[clip=true,width=.7\columnwidth,angle=90]{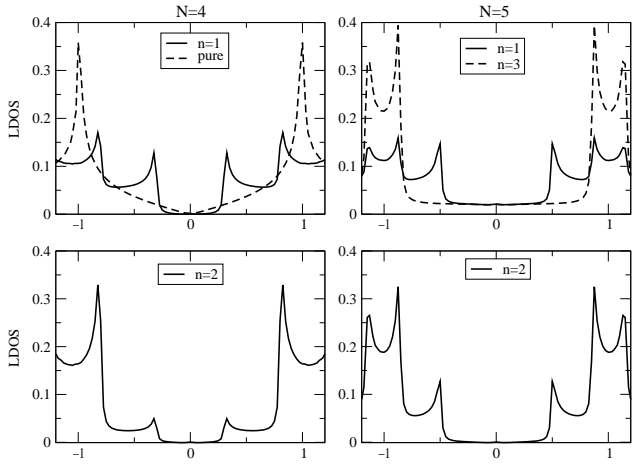}
\caption{Local density of states $\rho(x,\omega)$ vs.
$\omega/\Delta_{max}$ for a $(100)$ wire of width $N=4$ (left panels)
and $N=5$ (right panels), using $\Delta_0=0.2t$, $\mu=0$. }
\label{fig:wires100}
\end{center}
\end{figure}

Ziegler et al.\cite{zieglerwire}  investigated the problem of
(100) $d$-wave quantum wires in some detail, and discovered the
existence of a number-parity effect as a function of the width $N$
of a mesoscopic $d$-wave wire: a finite total DOS is found at the
Fermi level for odd $N$ and zero DOS (with a full gap in the
excitation spectrum, not a $d$-wave like gap) is found for even
$N$, at least for a simple tight-binding band at half-filling. The
differences between even and odd $N$ were shown to survive for
more general bands, as well.   For completeness, we reproduce,
using our approach, some of their LDOS spectra in Fig.
\ref{fig:wires100}.

\subsection{(110) wires}

\subsubsection{Normal metal wires}
\label{sec:nwires} 
\begin{figure}[tbh]
\begin{center}
\leavevmode
\includegraphics[width=.7\columnwidth]{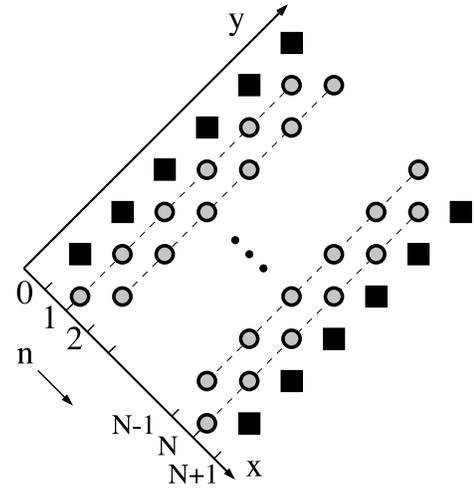}
\caption{Geometry of 110 wire (sites are filled circles) of width
$N$ bounded by 2 impurity lines (filled squares).  Wire is
infinite in both $y$ and $-y$ directions.} \label{fig:strip110}
\end{center}
\end{figure}

Based on intuition from the surface case, we
would expect that Andreev states play an important role  in (110)
wires, with geometry shown in Figure \ref{fig:strip110}. A crucial
question which arises in the following discussion is, how does one
identify a subgap state of true Andreev character? By merely
measuring the LDOS with an STM, for example, one may see several
peak structures, not all of which will be related to Andreev
reflections at the surfaces. One set of candidate states which
needs to be investigated first is the set of discrete dispersive
(with respect to $k_y$) levels which arise simply because of the
finite wire width.  These are of course present already in the
normal state wire.

The $k_x$-integrated bare normal state Green's function for an
infinite lattice above $T_c$ takes the form : \bea
{G}^{(0)}_{11}(n,k_y,\omega)&=&\frac{d}{2\pi}\int^{\pi/d}_{-
\pi/d}\frac{e^{ik_xnd}dk_x} {(\omega+i0)-\xi_\k}\nonumber\\
&=& - {i\exp [i|n| \arccos (-\omega/q(k_y))]\over
\sqrt{q^2(k_y)-\omega^2}}\, ,\eea  where $ q(k_y)$ is defined in
Eq.(\ref{qky}). The full Green's function for a (110) wire may
then be obtained by solving the $T$-matrix equation as
\be
G_{11}(n,n';k_y,\omega)=\frac{2\sin[n_{min}z]\sin[(n_{max}-N-
1)z]}
{\sqrt{q^2(k_y)-\omega^2}\sin[(N+1)z]}
\label{gnm}
\ee
where $n,n'=1,2,...,N$, $z\equiv z(k_y,\omega)=\cos^{-1}(-\omega/q(k_y))$,
$n_{min}={\rm min}(n,n')$,  and $n_{max}={\rm max}(n,n')$.

For every value of $k_y$, there generally exists a series of
eigenvalues which are the poles of the Green's function. They may
be obtained by simply solving
\begin{equation}
\sin[(N+1)z(k_y,\omega)]=0 \enspace ,
\label{eigennormal}
\end{equation}
which yields:
\begin{equation}
\omega_\nu(k_y)=-q(k_y)\cos\frac{\pi \nu}{N+1},~~
\nu=0,1,...,N+1 \ .
\label{modefreqnormal}
\end{equation}
This gives $N$ branches of solutions for $\nu = 1, ..., N$,
distributed symmetrically with respect to the Fermi level, and two
special solutions ($\nu=0$ and $N+1$) with $\omega=\pm q(k_y)$
which we refer to as defining the "effective band edge". One
interesting feature is the existence in the normal metal wire of
dispersionless zero-energy  quasiparticle states (ZES) which form
the branch $\nu=(N+1)/2$ when the width $N$ of the wire is odd.
Since the group velocity vanishes for dispersionless states, they
are always localized. We note that the case $N=1$ has only the
obvious dispersionless ZES since no transport is allowed with only
nearest neighbor hopping. In the general $N$=odd case, it appears
that there is always exactly one such localized state (doubly
degenerate in  particle-hole space) for any fixed $k_y$. A
deviation from half filling shifts the zero-energy states to
$-\mu$. Thus, for positive $\mu$ they are the hole states, while for
$\mu<0$ - electron states.

\begin{figure}[tbh]
\begin{center}
\leavevmode
\includegraphics[clip=true,width=.9\columnwidth]{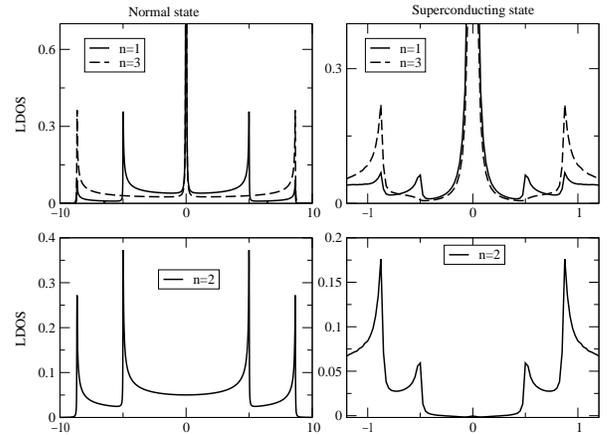}
\caption{Local density of states  for a (110)  wire of width
$N=5$, with $\Delta_0=0.2t$, $\mu=0$. Left: normal state,
$\rho(x,\omega)$ vs. $\omega/\Delta_{max}$; right: superconducting
state, $\rho(x,\omega)$ vs. $\omega/\Delta_{max}$.}
\label{fig:N5wires}
\end{center}
\end{figure}

The contribution of each of these states to the LDOS can be
estimated by examining the residue near the pole, where the
Green's function can be approximated as:
\begin{equation}
G(n,n^{\prime};k_y,\omega)
\approx\frac{Q_\nu(n,n^{\prime})}{\omega-\omega_\nu(k_y)},
\end{equation}
with the residue given by:
\begin{equation}
Q_\nu(n,n')=\frac{2\sin\frac{n\pi \nu}{N+1}\sin\frac{n^{\prime}\pi
\nu}{N+1}}{N+1},\label{weightn0}
\end{equation}
where $n$ is the index of the layers. Note from
(\ref{weightn0}) and the spectral representation of the Green's
function near a pole, it is easy to read off the quasiparticle
wavefunctions as $\psi_\nu(n) =\sqrt{2/(N+1)}\sin[n\pi\nu/(N+1)]$,
i.e. just the wave functions of a free particle confined to a box
of width $(N+1)d$.

\begin{figure}[tbh]
\begin{center}
\leavevmode
\includegraphics[clip,width=.9\columnwidth]{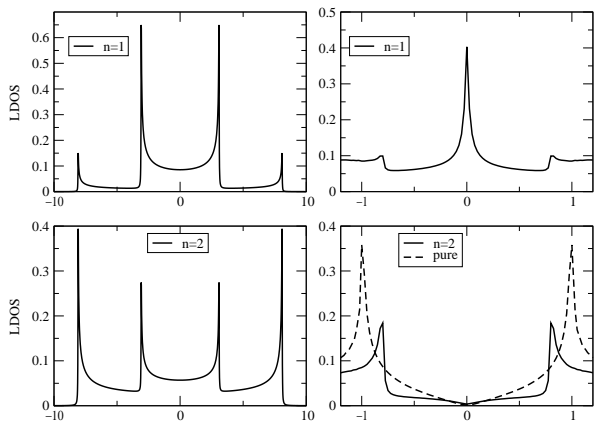}
\caption{LDOS for a (110)  wire of width $N=4$ with
$\Delta_0=0.2t$, $\mu=0$. Left: normal state, $\rho(x,\omega)$ vs.
$\omega/\Delta_{max}$; right: superconducting state,
$\rho(x,\omega)$ vs. $\omega/\Delta_{max}$.} \label{fig:N4wires}
\end{center}
\end{figure}

It is easy to check that for any $n$, $n'$ the residue $Q_\nu$
vanishes everywhere for the "states" at the effective bandwidth $\nu=0$ or
$N+1$; we therefore do not discuss them further, but focus on the
$N$ branches with finite residue. For these states the
residue can vanish locally on chains with number $n$, for
which the quantity $n\nu/(N+1)$ is an integer. The odd $N$ ZES,
for example, has a residue
\begin{equation}
Q_{\frac{(N+1)}{2}}(n,n)=\frac{2(\sin\frac{n\pi }{2})^2}{N+1}
\enspace .
\label{weightnn}
\end{equation}
As is seen from Eq.(\ref{weightnn}), the probability density of
the ZES oscillates with a period $2d$, taking finite values only
for odd $n$ and vanishing on all nearest neighbor sites (~where $n$
is even~). This is also valid for the states with energies $\pm\mu$
in the case of the deviation from half-filling and ensures no transport
with nearest neighbor hopping for quasiparticles with energies $\pm\mu$.

All these states are not surface bound states, as one can easily
check that their amplitude does not decay across the wire. The
quasiparticle spectrum is discrete (for fixed $k_y$) due to the
finite width of the wire and transforms into conventional
continuous quasiparticle spectrum in the massive normal metal in
the limit $N\to\infty$. The momentum resolved LDOS in the normal
metal wire with discrete dispersive states takes the form
\begin{equation}
\rho(n,k_y,\omega)=\sum\limits_{\nu=1}^{N}Q_{\nu}(n=n')\delta(\omega-
\omega_{\nu}(k_y)) \enspace .
\end{equation}
The integration over $k_y$ gives the LDOS:
\begin{equation}
\rho(n,\omega)=\frac{1}{2\pi d}\sum\limits_{\omega_{\nu}(k_y)=\omega}
\dfrac{Q_{\nu}(n,n)}{\left|\dfrac{d\omega_{\nu}}{dk_y}\right|} \enspace ,
\label{rhon}
\end{equation}
where the sum is taken over those $k_y$ and $\nu$, which satisfy
the equation $\omega_{\nu}(k_y)=\omega$.

The position of peaks in LDOS are then determined by extrema of
the dispersive branches (\ref{modefreqnormal}), taking place at
$k_y=0$: \be \omega_{\nu,peaks}=-4t\cos\frac{\pi \nu}{N+1},~~
\nu=1,...,N \ . \label{energypeaks}
\end{equation}

The peaks corresponding to the $N$ normal metal wire states are
seen clearly in Figs. \ref{fig:N5wires} and \ref{fig:N4wires} at
the eigenfrequencies given by Eq. (\ref{energypeaks}). It is easy
to check that the weights agree with Eq. (\ref{weightn0}). The
LDOS for the normal metal wires qualitatively differs from the
LDOS for bulk normal metals with the nearest neighbor hopping on
the square. The zero-energy peak (the Van Hove singularity) in the
bulk metal is symmetric as a function of the energy and its log
singularity is much broader than the  $\delta$-like peaks we find
here for wires.

\subsubsection{Superconducting wires}
\label{sec:swires}

The $T$-matrix and Green's function equations are necessarily more
complicated in the presence of superconductivity, but they are
still tractable in the $(110)$ case. The bare Green's functions
$\hat{G}^{(0)}$ are given by Eqs. (\ref{geven}) and (\ref{godd}),
as before. One must then solve Eq. (\ref{det}), which applies to
any situation which requires two lines of impurities, for the
eigenenergies $\omega$. Similar to the normal metal case, there are
special solutions of Eq. (\ref{det}): $\omega=\pm\Delta(k_y)$,
$\omega=\pm q(k_y)$ for any $N$ and $\omega=0$ for even $N$. They
are  poles of the $T$-matrix, but do not correspond to the poles
of the full Green's function.

The behavior of subgap surface states on the narrow
superconducting wire differs qualitatively from the case of the
superconducting halfspace due to the interference of the wave
functions of the states on both surfaces. Since the zero-energy
peak in the LDOS for the half-filled surface vanishes on each even
chain (~see Figs.(\ref{fig:surf110}), (\ref{fig:Andreev})), the
spectrum of  Andreev states on the $(110)$ wires becomes strongly
dependent on the parity of the number $N$ of chains in the
half-filled wire. This effect is quite pronounced for wires whose
width is the order of or less than the superconducting coherence
length. As we demonstrate below, some new qualitative features
arising in the superconducting state of the $(110)$ wires in the
quasiparticle spectrum above $\Delta(k_y)$ (~see Eq. (\ref{dky}))
can also be strongly dependent on the parity of $N$.

\vskip .2cm

{\it Odd $N$.} For a wire with odd $N$, only the Green's functions
(\ref{geven}) with even arguments  $n=0, \pm(N+1)$ enter
Eq.(\ref{det}).  Since at small frequencies these Green's function
are $\propto \omega$, it is straightforward to show that the only
subgap state is a dispersionless ZES $\omega=0$. Dispersive modes
exist as well and take the form

\begin{eqnarray} \omega_{\nu}^2(k_y)=q^2(k_y)\cos^2\frac{\pi
{\nu}}{N+1}+\Delta^2(k_y)\sin^2\frac{\pi
{\nu}}{N+1}~, \nonumber\\
~~{\nu}=1,...,\frac{N-1}{2} \ .
\label{dm}
\end{eqnarray}

This exactly coincides with the quasiparticle spectrum in a bulk
two-dimensional superconductor $\omega^2(k_y, k_x)=\xi^2(k_y, k_x)
+\Delta^2(k_y,k_x)$, if the discrete values of momentum component
across the wire $k_{x,\nu}=\pi{\nu}/(N+1)d$ are introduced. We
note that the dispersionless zero-energy Andreev states are the
only true subgap states in the spectrum, since the energy
$|\omega_{\nu}(k_y)|$ of the dispersive states (\ref{dm}) lie
above the respective value $|\Delta(k_y,k_x)|$ of the order
parameter, for any $\nu$ and $k_y$.

\begin{figure}[htb]
\begin{center}
\leavevmode
\includegraphics[width=.9\columnwidth]{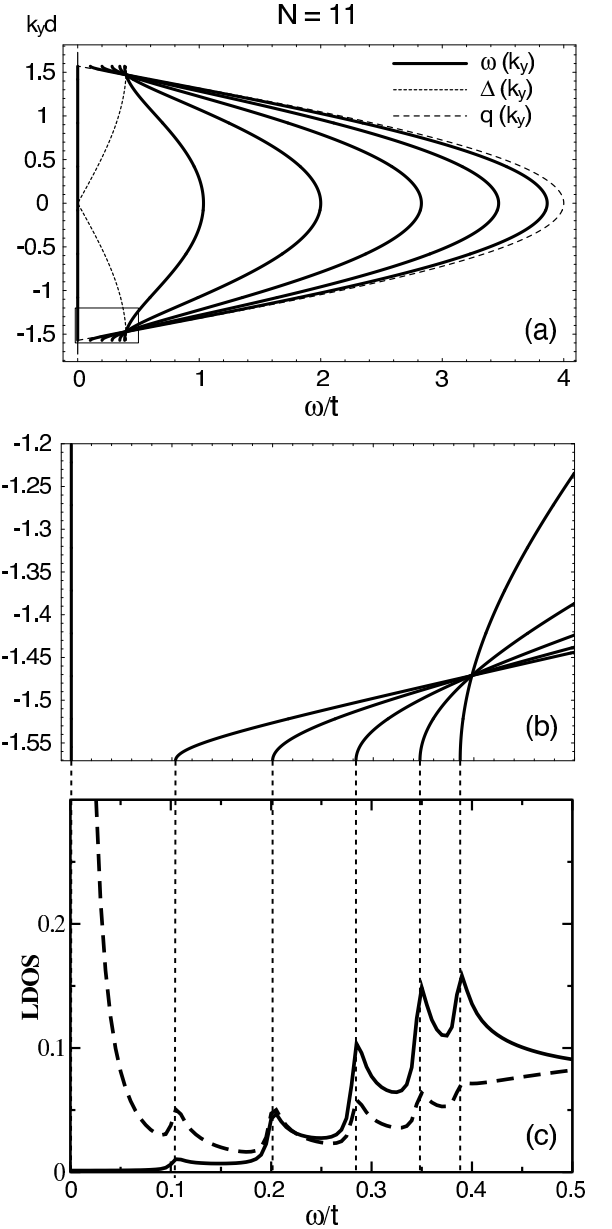}
\caption{ Quasiparticle spectra and the LDOS for superconducting
half-filled (110) wire with $N=11$.\ (a) Dispersive modes ($k_yd$
vs. $\omega/t$) for $N=11$, $\Delta_0=0.2t$, $\mu=0$. (b) The
blowup of the dispersive spectra near the edge of the Brillouin
zone.\ (c) The LDOS for energies less than $\Delta_{max}$: n=2
(solid line), n=1 (dashed line), where a broadening $\delta=.005t$
has been used. }
\label{figN11}
\end{center}
\end{figure}

There are, as in the normal state case, $N-1$ dispersive modes in
addition to the ZES. They are doubly degenerate due to the
particle-hole symmetry. This simple result can be understood as
follows. For a fixed $k_y$ the problem reduces to a
one-dimensional problem. The corresponding two-component
Bogoliubov-DeGennes wave function takes its values on $N$ sites,
resulting in $2N$ degrees of freedom in the system. With this
point of view it appears natural that for fixed $k_y$ the total
number of levels, which are twice degenerate, is $N$. The set of
levels with positive energies for $N=11$ is represented in panel
(a) of Fig.\ref{figN11}.

The LDOS for the superconducting wires with N=5 is shown in the
right panel of Fig.~\ref{fig:N5wires}. The dispersionless ZES
results in a pronounced peak at zero energy on odd layers.
Furthermore, each extremum of the dispersive mode $\omega(k_y)$
results in the peak in the LDOS at the energy
$\omega_{k_{y,extr}}$. One series of  peaks is associated with the
extrema at $k_y=0$. Since $\Delta(k_y=0)=0$, the peaks lie at the
same positions as in the normal metal wire (\ref{energypeaks}).
Although they are irrelevant to the superconducting properties of
the wire, some of these peaks can lie at finite energies below the
maximum of the gap function $\Delta_{max}=2\Delta_0$. For
instance, the lowest position at finite energies of the peaks of
this series is $4t\sin\frac{\pi}{N+1}$. This can be both above or
below $\Delta_{max}$, depending on the ratio $\Delta_{max}/t$ and
the wire width $N$. In contrast with the normal metal wires, the
dispersive quasiparticle modes Eq.(\ref{dm}) in the
superconducting wires have extrema also at the edge of the
surface-adapted Brillouin zone $k_y=\pm \pi/(2d)$ \cite{ebz}. They
are shown in panel (b) of Fig.\ref{figN11} for the wire with
$N=11$. This leads to additional series of $(N-1)$ quasiparticle
peaks in the LDOS for the superconducting wires, which lie below
$\Delta_{max}$:

\begin{equation} \omega_{peaks}=\pm \Delta_{max}\sin\frac{\pi \nu}{N+1},~~
\nu=1,...,\frac{N-1}{2} \ .
\label{energypeaksodd}
\end{equation}
Panel (c) of Fig.\ref{figN11} displays the series of peaks in the LDOS for
$N=11$.

The  dispersive states forming the nonzero low-energy peaks
(\ref{energypeaksodd}) in LDOS, are {\it not} Andreev states. They lie
below $\Delta(k_y)$ for $k_y$ near the edges of Brillouin zone, but they
are situated above the bulk gap function $\Delta(k_x(\nu),k_y)$. The wave
function for any of these states possesses the finite current of
the probability density, while for Andreev states the total
probability current vanishes.

While positions of the peaks are determined by the
extrema of the dispersive energies, their weights in the LDOS
are controlled also by quasiparticle wave functions, which form the
quantity $Q(n,n)$ in Eq.(\ref{rhon}).
For this reason the weights of the peaks can substantially
differ on different layers and may vanish on some of them, quite
analogously to the normal metal wires (see Eq.(\ref{weightn0})).
Due to an interference from two surfaces, the wave function, taken
on even layers, turns out to coincide with the respective wave
function in the normal metal case (apart from a normalization
constant). On the other hand, the wave function on odd layers is a
superposition of electron-like and hole-like Bogoliubov
quasiparticles on the wire.

\vskip .5cm {\it Even $N$.}
In the case of even $N$ half-filled wires, the quasiparticle
spectra become more complicated. The Green's functions
(\ref{geven}), (\ref{godd}) with even ($n=0$) and odd
($n=\pm(N+1)$) arguments enter Eq.(\ref{det}), which can be
reduced to  the following form
\be q(k_y)\tan[(N+1) z]=
\alpha\Delta(k_y)\tan z \enspace ,
\label{z}
\ee
where $\alpha=\pm
1$, $0\le z\le\pi/2$. Solutions $z_{\nu,\alpha}(k_y)$ of
Eq.\eqref{z} are directly associated with $\omega$, in accordance
with Eq.\eqref{zomega}:
\begin{equation}
\omega^2_{\nu,\alpha}(k_y)=q^2(k_y)\cos^2z_{\nu,\alpha}(k_y)+
\Delta^2(k_y)\sin^2z_{\nu,\alpha}(k_y) \enspace .
\label{evenspectr}
\end{equation}
Here ${\nu}=1,...,\frac{N}{2}$ and $\alpha$ are the indices of the
solution. Comparing Eqs.\eqref{evenspectr}, \eqref{dm} for the
spectra of odd and even wires, one can see that $z/d$ plays the
role of effective discrete values of the momentum component $k_x$
(at fixed $k_y$) across the wire. Eq.\eqref{z} can be transformed
to a polynomial equation in $\tan z$ of the $N$-th degree, if one
excludes the special solutions $\omega=\pm\Delta(k_y), \pm q(k_y)$
mentioned above. Hence, for a fixed $\alpha$ there are exactly
$N/2$ positive and $N/2$ negative solutions for $\omega$,
describing $N$ dispersive branches of the quasiparticle spectrum.
Explicit analytical form of the spectra can be easily found from
Eqs.\eqref{evenspectr}, \eqref{z} in the particular cases $N=2$
\be
\omega_{\alpha}=\pm\dfrac{1}{2}\Bigl(q(k_y)+\alpha\Delta(k_y)\Bigr),
\label{2} \ee and $N=4$ ($\nu=\pm$, $\alpha=\pm$)
\begin{eqnarray}
&\omega^2_{\nu,\alpha}(k_y)=\Delta^2(k_y)+\dfrac{1}{8}q(k_y)\Bigl(q(k_y)+
\alpha\Delta(k_y)\Bigr)\Biggl[3- \nonumber \\
& -5\alpha\dfrac{\Delta(k_y)}{q(k_y)}
+\nu\,\sqrt{5-6\alpha\dfrac{\Delta(k_y)}{q(k_y)}+
5\dfrac{\Delta^2(k_y)}{q^2(k_y)}}\,\Biggr] . \label{4}
\end{eqnarray}
Eq.\eqref{4} is defined for all $k_y$ in the Brillouin zone for
which $\omega^2_{\nu,\alpha}(k_y)$ is positive. Eqs.\eqref{2},
\eqref{4} describe the quasiparticle spectra for
non-self-consistent wires with small numbers of chains. As we will
show in Sec. \ref{sec:self}, the self-consistent treatment of the
problem can lead to important modifications of the results, at
least if the width of the wire is less than or of  order  the
superconducting coherence length.

\begin{figure}[tbh]
\begin{center}
\leavevmode
\includegraphics[width=.9\columnwidth]{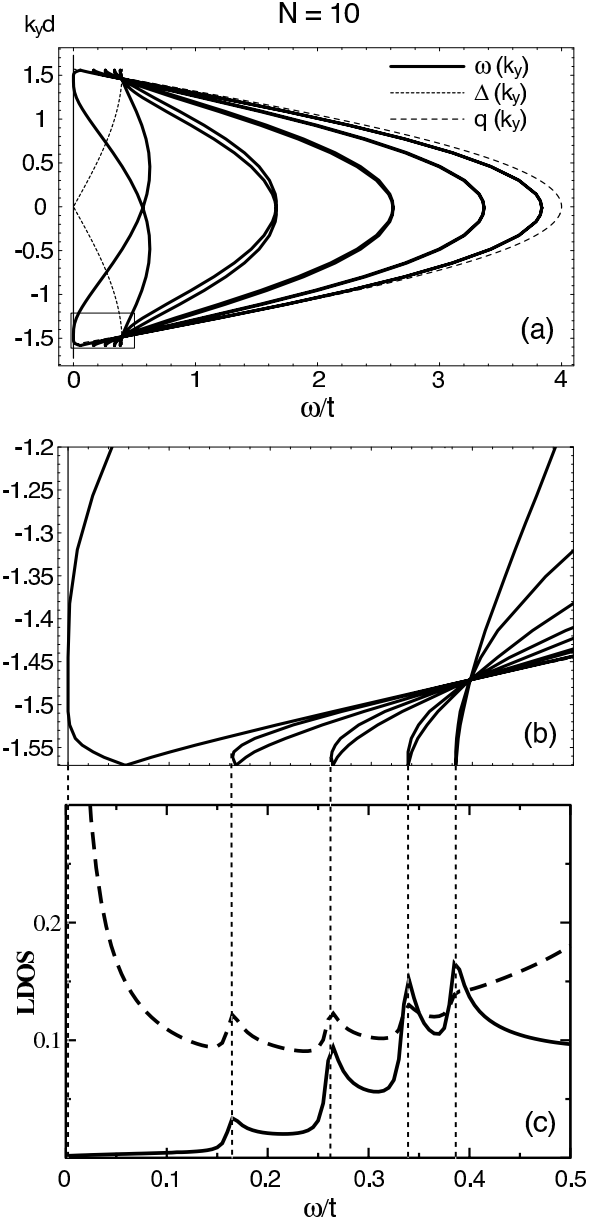}
\caption{Quasiparticle spectra and the LDOS for superconducting
half-filled (110) wire with $N=10$.\ (a) Dispersive modes ($k_yd$
vs. $\omega/t$) for $N=10$, $\Delta_0=0.2t$, $\mu=0$.\ (b) The
blowup of the spectra  near the edge of the Brillouin zone.\ (c)
The LDOS for energies less or equal to $\Delta_{max}$ : n=2 (solid
line), n=1 (dashed line), where a broadening $\delta=.005t$
has been used. }
\label{figN10}
\end{center}
\end{figure}

Whereas in the normal metal state, $\nu$ is the only index of the
solution (for a given $k_y$), in the superconducting even-$N$ wire
each branch with fixed $\nu$ splits into two, corresponding to two
values of the index $\alpha$, giving a  total  of $2N$ distinct
branches. Since the order parameter vanishes for $k_y=0$, the
splitting is absent in this particular case. The splitting is
associated with the symmetry breaking of the eigenstates with
respect to the sign reversal of $k_y$. In the normal  even-$N$
wires, as well as in the odd $N$ wires (both in the normal metal
and in the superconducting states), each separate branch of the
spectra $\omega_{\nu}(k_y)$ is an even function of $k_y$. This is
not the case, however, for the even $N$ superconducting $(110)$
wires. As directly seen from Eq.\eqref{z} and the relations
$q(-k_y)=q(k_y)$, $\Delta(-k_y)=-\Delta(k_y)$, the solutions of
Eq.\eqref{z} $z_{\nu,\alpha}(k_y)$ with fixed $\alpha$ are not odd
or even functions of $k_y$, due to the nonzero right hand side of
Eq.\eqref{z} in the superconducting state. Evidently, each
separate branch of the spectra $\omega_{\nu,\alpha}(k_y)$ possesses
the same property. The whole spectrum, however, remains symmetric
with respect to $k_y\to-k_y$, since the branches with positive and
negative $\alpha$ simply interchange with each other under this
transformation. Spectra of the half-filled $N=10$ superconducting wire
with positive energies are shown in panel (a) of
Fig.~\ref{figN10}. They agree with the above discussion. For
branches with higher energies, the splitting is less than for lower
branches. For small splittings, $\delta\omega\propto \Delta/(N+1)$.

The splitting manifests itself by slightly shifting the positions
of peaks in the LDOS, due to respective shifts of extrema of
dispersive quasiparticle energies. In contrast with the double
number of  extrema, the total number of the peaks in the LDOS does
not change. Any peak in the LDOS originates from the two extrema,
situated symmetrically with respect to the sign of $k_y$. Each
pair of extrema belongs to two spectral branches, which differ
only by their index $\alpha=\pm1$. If we disregard the splitting
of the quasiparticle spectra, the positions of the peaks in the
LDOS coming, for example, from the edge of the Brillouin zone, are
as follows
\begin{equation} \omega_{peaks}=\pm \Delta_{max}\cos\frac{\pi \nu}{N+1},~~
\nu=1,...,\frac{N}{2} \ .
\label{energypeakseven}
\end{equation}
The splitting shifts the peak positions towards slightly lower
energies $|\omega|$, as compared with those in Eq.\eqref{energypeakseven}.
The respective extrema of dispersive energies are shifted towards lower
values of $|k_y|$ from $k_y=\pm\pi/(2d)$. These low-energy peaks in the
LDOS are shown in panel (c) of Fig.~\ref{figN10}.

The other peaks in the LDOS come, neglecting the splitting, from
the center of the Brillouin zone. Since the order parameter
vanishes in the center of the zone, the positions of the peaks are
still the same as in Eq.\eqref{energypeaks} for the normal metal
state. The splitting, taking place in the superconducting state,
slightly shifts the extrema of the spectra from the center of the
zone, so that the peaks move to a larger energy values $|\omega|$,
as compared with those in Eq.\eqref{energypeaks}.

The LDOS for the superconducting wire with $N=4$ is shown in the
right panel of Fig.\ref{fig:N4wires}. In addition to the peaks at
finite energies, discussed above, there is also a well pronounced
zero-energy peak there (see also panel (c) of Fig.~\ref{figN10}).
This peak originates from the extrema of two dispersive branches
of Andreev states, which take place at $\pm k_{y,0}d=\pm
\tan^{-1}(2t/\Delta_0)$, where the relation
$q(k_{y,0})=\Delta(k_{y,0})$ holds. Indeed, as  follows from
Eqs.(\ref{z}), (\ref{zomega}) and, in particular, directly seen
from Eq.(\ref{4}), the energy and its derivative equal zero for
$k_y=\pm k_{y,0}$. It turns out that the multiplicity of zeroes
of the lowest dispersive branches of states at the extrema $k_y=\pm k_{y,0}$ is $N/2$,
i.e. $\omega\propto |k_y-k_{y,0}|^{N/2}$ in the close vicinity of $k_{y,0}$.
This follows directly, for instance, from Eq.(\ref{omegaa}) given below.
We notice, that the lowest dispersive curve in panel (b) of Fig.~\ref{figN10}
manifests very slow change of energy when it touches the zero-energy line.
This agrees with our expectation $\omega\propto |k_y-k_{y,0}|^{5}$
for the (110) wire with $N=10$. The respective peak in the LDOS of wires with
large even $N$ diverges at $\omega=0$, but is not a delta-like peak as it is
for wires with odd $N$.

The identification of the Andreev nature of the quasiparticle bound
states in confined geometries turns out to be a nontrivial
problem, at least in the case of half-filled (110) wires with even
$N$, where there is a symmetry breaking with respect to
$k_y\to-k_y$. As seen from Eqs.(\ref{z})-(\ref{4}), the asymmetry
can be associated with the order parameter behavior
$\Delta(-k_y)=-\Delta(k_y)$ and with the sensitivity of the
quasiparticle energies to the $\pi$-shift of the order parameter
phase. A dependence of the dispersive energy curve on the order parameter
phase is usually an intrinsic feature of Andreev bound states only.
We consider the vanishing of  the probability density current as an
defining property of  Andreev bound states. These states should
be able to carry finite electric current, however. In odd $N$ wires
only zero-energy dispersionless quasiparticle states satisfy the
above requirements. In even $N$ $(110)$ wires the zero-energy
dispersionless states do not arise at all. The interference of
wave functions located near two surfaces of the even $N$ wire
induces dispersive Andreev branches $\pm\omega_A(k_y)$, which
transform into the zero-energy states in the limit of infinitely
large $N$. Moreover, all quasiparticle states in the half-filled
$(110)$ superconducting wire with even $N$, even those lying above
the gap, turn out to satisfy the above mentioned conditions, i.e.
possess the properties of Andreev bound states. As an example, we
describe below in detail the structure of the lowest dispersive
quasiparticle branch, which we denote as Andreev branch
$\omega_A(k_y)$, although its energy varies with $k_y$  over a
large range, including both the subgap and the supergap regions
(see, for example, Fig.~\ref{figN10}).

For those $k_y$ where ${\rm
min}\{|\Delta(k_y)|,q(k_y)\}<|\omega_A(k_y)| <{\rm
max}\{|\Delta(k_y)|,q(k_y)\}$, the wave function of the state can
be written on odd layers ($n=2m+1$) as: \be \left(\!
\begin{array}{c} u\\v \end{array}\!\right)= C{\rm sgn}\omega_A
\sin[(\!N\!+\!1\!-\!n\!)z] {1\choose {i{\rm sgn}k_y}}
\label{aodd}
\ee
and on even layers $n=2m$
\be
\left(
\begin{array}{c} u\\v\end{array}\right)= C(-1)^{N/2}\sin(nz)
{1\choose {-i {\rm sgn}k_y}} \enspace .
\label{aeven}
\ee

In the range of $k_y$ for which
$|\omega_A(k_y)|<{\rm min}\{|\Delta(k_y)|, q(k_y)\}$,
the quantity $z$, entering Eq.\eqref{evenspectr}, becomes imaginary.
Under the condition $|\omega_A(k_y)|<|\Delta(k_y)|<q(k_y)$
the wave function takes the following form on odd layers
($n=2m+1$):
\be
\left(\! \begin{array}{c} u\\v \end{array}\!\right)=
C_1 (-1)^m {\rm sgn}\omega_A
\sinh[(\!N\!+\!1\!-\!n\!)z_1] {1\choose {i{\rm sgn}k_y}}
\label{aodd1}
\ee
and on even layers $n=2m$
\be
\left( \begin{array}{c} u\\v\end{array}\right)=
C_1(-1)^m\sinh(nz_1) {1\choose {-i {\rm sgn}k_y}} \enspace .
\label{aeven1}
\ee
Analogously, if $|\omega_A(k_y)|<q(k_y)<|\Delta(k_y)|$, the wave function
on odd layers ($n=2m+1$)
\be
\left(\! \begin{array}{c} u\\v \end{array}\!\right)=
C_2 {\rm sgn}\omega_A
\sinh[(\!N\!+\!1\!-\!n\!)z_1] {1\choose {i{\rm sgn}k_y}}
\label{aodd2}
\ee
and on even layers $n=2m$
\be
\left( \begin{array}{c} u\\v\end{array}\right)=
C_2 \sinh(nz_1) {1\choose {-i {\rm sgn}k_y}} \enspace .
\label{aeven2}
\ee
Here $C$, $C_1$ and $C_2$ are normalization constants,
$z_1=|{\rm Im}z(\omega_A(k_y),k_y)|$, $z$ is defined in
Eq.\eqref{zomega} and taken at $\omega=\omega_A(k_y)$.

The condition $|u(n,k_y)|=|v(n,k_y)|$, which is valid for all
solutions Eqs.~\eqref{aodd} -- \eqref{aeven2},  results in
zero total probability current density, while the electric current
does not vanish for the given branch. This ensures the Andreev
character, as defined above, of all the states in even $N$ wires,
regardless of whether their energies lie above or below the gap.

The quasiparticle states which belong to the same dispersive
branch $\omega_A(k_y)$ can have their wavefunctions both
symmetrically decaying in the depth of the wire
(Eqs.~\eqref{aodd1} -- \eqref{aeven2}) or oscillating and forming
standing waves across the wire (Eqs.~\eqref{aodd}, \eqref{aeven}).
In particular, the amplitude of the wave functions in
Eqs.~\eqref{aodd1} -- \eqref{aeven2} takes its maximum value on
layers $n=1$ and $n= N$, and manifests  Friedel-like oscillations,
as the layer index changes from odd to a neighbor even value or
vice versa. On a larger scale the amplitude decays in the bulk of
the wire symmetrically with respect to two surfaces.

It is instructive to follow how the above results transform into
the well known dispersionless zero-energy Andreev surface states
in the limit of large $N$. For sufficiently large width of the
wire compared with the coherence length, and for those $k_y$ where
the wavefunction decays in the depth of the wire, the dispersive
energy $\omega_A(k_y)$ takes a relatively simple form:
\begin{eqnarray}
\omega_A(k_y)=\pm\dfrac{2\Delta(k_y)
q(k_y)}{\sqrt{|q^2(k_y)-\Delta^2(k_y)|}}\times
\qquad \qquad \qquad \quad \nonumber\\
\times\exp\left[-(N+1){\sinh}^{-1}\dfrac{{\rm min}\{|\Delta(k_y)|,
q(k_y)\}}{\sqrt{|q^2(k_y)-\Delta^2(k_y)|}}\right] . \label{omegaa}
\end{eqnarray}
It follows from Eq.\eqref{omegaa}, that the energy of the Andreev
states is exponentially small and vanishes in the limit of
infinitely large $N$. With decreasing energy, the range of $k_y$
where the wave function oscillates (see Eqs. \eqref{aodd},
\eqref{aeven}), converges to the center and to the edges of the
Brillouin zone and finally collapses to the respective points.
Hence, in the limit of very large $N$ the amplitude of the wave
function decays inside the wire for practically all values of
$k_y$. This means that for $N\to\infty$ the dispersive branch
$\omega_A(k_y)$ transforms into the zero-energy dispersionless
surface states situated near the two surfaces.

\begin{figure}[tbh]
\begin{center}
\leavevmode
\includegraphics[clip=true,width=.9\columnwidth]{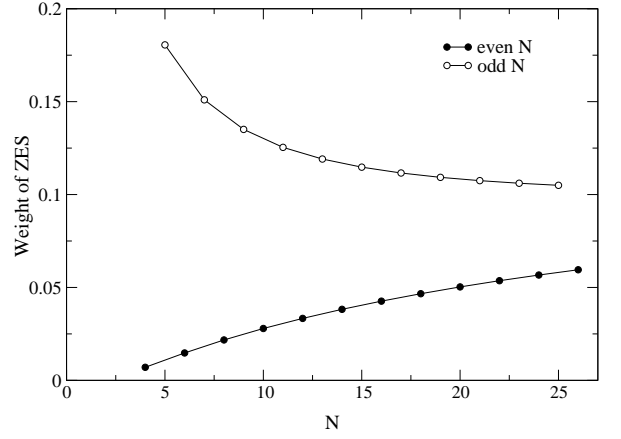}
\caption{The weight of the zero-energy peak in the LDOS, taken at the first layer,
as a function on odd (open circles) and even (filled circles) $N$.}
\label{figweights}
\end{center}
\end{figure}

Fig.~\ref{figweights} shows $N$-dependence of the weight of the
zero-energy peak in the LDOS. For odd $N$, the weight diminishes
with increasing $N$, while for even $N$ it increases. In the limit of
large $N$ the two curves will converge to the weight for the
zero-energy surface states when the surfaces are infinitely far
apart. The odd-even effects become negligibly small only for
$Nd\gg \xi_0$. One could try to recover the quasiclassical
results for wide wires $Nd\gg a$ after averaging over odd and even
film widths. The boundary conditions for
quasiclassical propagators are taken somewhere on a distance $l$
from the surface and imply some uncertainty about the boundary
positions, as well as the film (wire) thickness ($a\ll l\ll \xi_0$).
As seen from Fig.~\ref{figweights}, averaging of the weights of the
peaks over odd and even $N$ will strongly reduce their width
dependence. This kind of averaging is much closer (although not
identical) to the quasiclassical results on the width dependence of
the LDOS for the $d$-wave superconducting film \cite{nagai}.  Very
recent quasiclassical results \cite{fominov}, treating various wire
orientations, demonstrate the appearance of energy bands of
quasiparticle states, in particular, for $(110)$ wires. Our
microscopic model for high quality half-filled wires with fixed
number of chains gives, however, only a couple of branches
Eq.\eqref{omegaa} for even $N$ wires and the dispersionless
zero-energy states for odd
$N$ wires. We associate the difference between the microscopic and
the quasiclassical results with the particular condition of
half-filling. The quasiclassical approach implies no singular
behavior of the LDOS in the normal metal state near the Fermi
surface, whereas the Van Hove singularity takes place in the LDOS
on the Fermi surface for the normal metal state of half-filled
infinite square lattice. An agreement of our microscopic results with
the quasiclassical ones arises in the presence of deviations from
half-filling (see below Eqs.\eqref{omegaamu}),\eqref{phi}).

As already mentioned, in the even $N$ wires the particle-hole
structure of any quasiparticle states with broken $k_y\to-k_y$
symmetry satisfies the condition $|u(n,k_y)|=|v(n,k_y)|$. Our
picture is that for the states above the gap this Andreev
particle-hole structure is generated by the infinite sequence of
``overbarrier'' (overgapped) Andreev reflections, induced by a
sign reversal of the order parameter, which the quasiparticles
experience along their trajectories being bounded inside the wire
with impenetrable surfaces. This unconventional feature does not
take place for the states above the gap in the odd $N$ wires,
since the two surfaces always result in the standing waves across
the wire with no important interference effects in this case. For
negligibly small splitting one should consider a superposition of
two wave functions, describing the two split states. Then the
Andreev structure of initially nondegenerate wave functions is
lost, since the moduli of particle and hole amplitudes can easily
differ from each other. The Andreev structure of quasiparticle
wavefunctions can be lost also in the presence of deviations from
the half-filling, if $\mu$ is larger or of the same order as the
splitting. Since the splitting vanishes for $k_y=0$, one can
expect that for sufficiently small $k_y$ the Andreev structure of
the wave functions will be destroyed even for small $\mu$. In the
next section some other consequences of deviations from
half-filling are considered.

\subsubsection{Deviations from  half-filling}

The shape of the Fermi surface depends on $\mu$ and has a strong
influence on the low-energy quasiparticle spectrum. Consider, for
example, low-energy quasiparticle states under the conditions
$|\omega|,\ |\Delta(k_y)|\ll q(k_y)$. This ensures that the
quasiparticle energies lie  close to the Fermi surface and one can
take their energies in the normal metal state to be in the linear
form $\propto \v_F\cdot(\k-\k_{F})$. Under this approximation,
effects of the particle-hole asymmetry are small and one can use a
quasiclassical approximation, which is valid for quasiparticles
close to the Fermi surface. Then only the order parameter on the
Fermi surface $\Delta(k_{F,x}, k_{F,y})$ enters the equations. For
the wire geometry the momentum component $k_{F,y}$ is an
independent parameter, while $k_{F,x}(k_{F,y})$, and
$\Delta(k_{F,x}(k_{F,y}), k_{F,y})$ are actually functions of
$k_{F,y}$. For $\mu=0$ the Fermi surface for the $(110)$ wire is a
square with sides parallel to $x$ or $y$ axes. Hence, $k_{F,x}=\pm
\pi/(2d)$ actually does not depend on $k_{F,y}$ in this case and
$\Delta(\pm\pi/(2d), k_{F,y})=\pm\Delta(k_{F,y})$, in accordance
with Eq.\eqref{dky}. For finite $\mu$ we find
$\Delta(k_{F,x},k_{F,y})=
\dfrac{\Delta(k_{F,y})}{q(k_{F,y})}\sqrt{q^2(k_{F,y})-\mu^2}$.

\begin{figure}[!tbh]
\begin{center}
\leavevmode
\includegraphics[clip=true,width=.9\columnwidth]{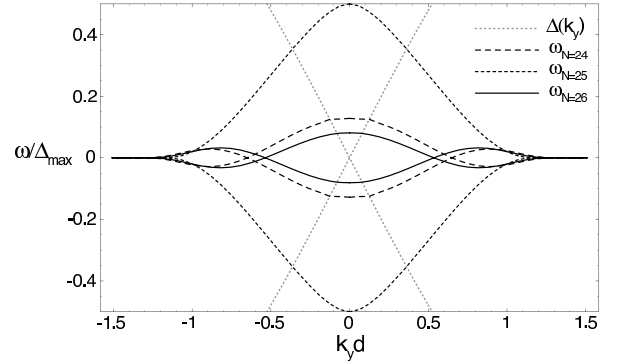}
\caption{The lowest energy branches for N=24, 25, 26. The parameters
are $t=2.5$, $\mu=0.2t=0.5$, $\Delta_0=0.2t=0.5$, $\Delta_{max}=2\Delta_0=1$.}
\label{fig:mu02spectrum}
\end{center}
\end{figure}

In the case of superconducting wires the equation for
quasiparticle subgap energies near the Fermi surface
$|\omega|<|\Delta(k_{F,x},k_{F,y})|$ takes the form
\begin{eqnarray} \omega^2\left(
\sinh^2\left[(N+1) d\dfrac{\sqrt{\Delta^2(k_{F,x},
k_{F,y})-\omega^2}}{
|v_{x,f}(k_{F,y})|}\right] \right.\nonumber\\
+\sin^2\phi(k_{F,y})\Biggr)= \Delta^2(k_{F,x},
k_{F,y})\sin^2\phi(k_{F,y})~,\label{spectrumassym}
\end{eqnarray}
where $\phi(k_{F,y}) \equiv k_{F,x}d(N+1)$.
 The lowest branches of quasiparticle spectra, which follow
from Eq. \eqref{spectrumassym} for $N=24,25,26$, are shown in
Fig.\ref{fig:mu02spectrum}. The solution of
Eq.~\eqref{spectrumassym} reduces to a simple form in the limit of
large $N$:

\bea \omega_A(k_{F,y})=\pm 2\Delta(k_{F,y})\sin\phi(k_{F,y})\times
\qquad \qquad \qquad
\nonumber\\
\exp\left[-(N+1)d\dfrac{|\Delta(k_{F,y})|}{|v_{x,f}(k_{F,y})|}\right]
\enspace . \label{omegaamu} \eea

For half-filled wires, when $\mu=0$, the phase $\phi(k_{F,y})$
does not depend on $k_{F,y}$, being equal to $\phi_{\rm odd}=m\pi$
for odd $N$ wires and $\phi_{\rm even}=(m+1/2)\pi$ for even $N$
wires. This difference between the phases $\phi_{\rm odd}$ and
$\phi_{\rm even}$ plays an important role in forming well
pronounced odd-even effects in the spectra of wires with odd and
even numbers of layers. Indeed, for half-filled odd $N$ wires
Eq.~\eqref{omegaamu} reduces to the zero-energy dispersionless
Andreev states. At the same time, for half-filled even $N$ wires
Eq.~\eqref{omegaamu} describes a dispersive branch of
quasiparticle energies, which coincides with Eq.~\eqref{omegaa}
under the condition $|\Delta(k_y)|\ll q(k_y)$.

\begin{figure}[t]
\begin{center}
\leavevmode
\includegraphics[clip=true,width=.9\columnwidth]{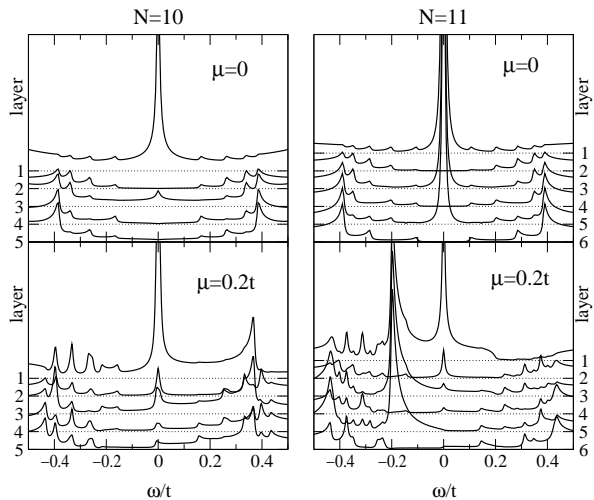}
\caption{Local density of states for a (110) wire
with N=10 (left column) and N=11 (right column); $\mu=0$ (upper panels),
$\mu=0.2t$ (lower panels). On each panel the different curves represent
various chains.}
\label{fig:mu110}
\end{center}
\end{figure}

For $\mu\ne0$ the phase $\phi(k_{F,y})$ noticeably depends on
$k_{F,y}$: \be \phi(k_{F,y})=k_{F,x}d(N+1)=
(N+1)\cos^{-1}\left(-\dfrac{\mu}{q(k_{F,y})}\right) \enspace .
\label{phi}\ee
Qualitative deviations of low-energy quasiparticle spectra,
shown in Fig.~\ref{fig:mu02spectrum}, from the respective
spectra of half-filled wires (see
Figs.~\ref{figN11},~\ref{figN10}) are associated with the behavior
of the phase $\phi(k_y)$.
The odd-even effect in the spectra of $(110)$ wires
becomes less pronounced in the case of finite $\mu$, as found by
Ziegler et al. \cite{zieglerwire}. For some values of $\mu$ the
spectra of odd $N$ and even $N$ wires may have no qualitative
differences at all. As it follows from Eq.~\eqref{omegaamu}
and, in a more general case, from Eq.~\eqref{spectrumassym},
for $\mu\ne 0$ additional and  strong dispersion of the spectra
comes from the $k_y$-dependence of the phase $\phi(k_{F,y})$. The
larger the $N$, more oscillations of $\sin\phi(k_y)$ take place
with varying the $k_{F,y}$. Hence, more extrema of
$\omega_A(k_{F,y})$ arise. This results in additional peaks in the
LDOS, which appear only in the presence of finite $\mu$.
As seen from Eq.~\eqref{phi}, the phase $\phi$ can considerably vary,
when the film (wire) thickness  varies from $Nd$ to $Nd-l$ and
$a\ll l\ll\xi_0$, $l\ll Nd$. Thus, in averaging the spectrum over the film
thickness, a large number of respective
additional peaks arises, filling the whole low-energy band
in the $(110)$ wire with edges described by Eq.~\eqref{omegaamu}. This
is in agreement with the quasiclassical results \cite{nagai,fominov}.

The particle-hole asymmetry is another important feature of the
spectra. It can be well pronounced for finite $\mu$, but lies
beyond the quasiclassical approximation. Fig.\ref{fig:mu110}
displays the asymmetric LDOS, calculated with Eq.(\ref{11}) for
$(110)$ superconducting wires with $N=10$ and $N=11$ in the case
$\mu=0.2t$. For  comparison, the respective LDOS for $\mu=0$ is
also shown. A well-pronounced peak at $\omega=-\mu$ arises with a
deviation from half-filling in the LDOS for the odd $N$ wires. We
remind the reader that in the half-filled normal state $(110)$
wires with odd $N$ the dispersionless zero-energy quasiparticle
states have been found in Sec.\ref{sec:nwires} (see
Eq.(\ref{modefreqnormal}) with $\nu=(N+1)/2$). The wave function
of these states, as well as the residue of the pole-like term in
the Green's function Eq.(\ref{weightnn}), is a standing wave
across the wire, taking zero values on alternating sites.
In the superconducting state the zero-energy standing wave
disappears and the dispersionless zero-energy Andreev surface
states arise. Their weight exponentially decays in the bulk of
wide odd $N$ wires. In the presence of a deviation from
half-filling the energy of the quasiparticle states in the normal
state odd $N$ wires shifts to $-\mu$. These dispersionless states
with finite energy keep the character of standing waves. Further,
in the superconducting wires with finite $\mu$ the low-energy
states become dispersive both for odd and even $N$
wires. As seen from Fig.(\ref{fig:mu02spectrum})
for the wire with $N=25$, the branch with lowest energy has in
this case two extrema. The maximal value of $|\omega|$ for the
states forming this branch, lies at $k_y=0$ and contributes to the
peak at $\omega=-\mu$ associated with the hole contribution, if
$\mu>0$. Since the order parameter vanishes at $k_y=0$, the
respective quasiparticle wave function is a standing wave and the
peak position coincides with that in the normal metal state of the
wire. The minimal value of $|\omega|$ is zero and contributes to
the zero-energy peak. The zero-energy peak is associated with
comparatively large value of $k_y$, comparable with the size of
the Brillouin zone, and the order parameter at this value of $k_y$
is of order  $\Delta_{max}$. Thus, the zero-energy peak is
associated with surface Andreev states, whose quasiparticle
wave function decays in the bulk of wide odd $N$ wires. For narrow
wires, the self-consistency condition becomes important. As shown
in the next section, at finite $\mu$ the self-consistency
condition can lead to more important consequences as compared with
the case $\mu=0$.

\label{sec:finitemu}

\section{Self-consistent treatment of $\mathbf{(110)}$-wires}
\label{sec:self}

In previous sections, the order parameter was assumed constant
over the whole width of the wire in order to allow for analytical
solutions. Even for a single $(110)$-surface, however, a
self-consistent treatment of the order parameter gives rise to
interesting effects: the $d_{x^2-y^2}$--wave order parameter is
strongly suppressed near the surface and a complex $is$--admixture
(or some other time-reversal symmetry-breaking state) is possible, which
leads to a splitting of the zero-energy Andreev bound state~\cite{mats95,%
buch95,cov97,fog97,Ting110,deutsch99,greene03,kos01,sigrist,ohashi99,%
vanHarlingen02,deutsch02,deutsch03}.
Even larger effects are therefore to be expected for the wire limited
by two $(110)$-surfaces.
Indeed, our self-consistent evaluation indicates that for very
narrow wires a quasi one-dimensional triplet superconducting state
can replace the conventional $d_{x^2-y^2}+is$--state. For finite
chemical potential
$\mu$ the normal metal state can become energetically favorable
as the ground state for narrow wires, while superconductivity recovers
with increasing wire width. Under special
conditions, even a mixture of singlet- and triplet-pairing can
occur.
It is important to recall at this point that one
expects mean field theory to break down as the 1D limit is
approached even at $T=0$.   Thus the predictions for various kinds
of superconducting order mentioned below are to be treated with
some skepticism as regards quantitative predictions.  Nevertheless
we view our results as presenting intriguing evidence that when
surface energies begin to become comparable to the energy
differences between $d$--wave and other bulk pair channels, strong
fluctuations with symmetries optimal for quasi-1D system,
including spin triplet pair fluctuations, will result.

\subsection{Order parameter}

Self-consistent solutions to Hamiltonian~(\ref{ham}) were
obtained by solving the Bogoliubov-de Gennes equations for
$(110)$-wires,
\begin{eqnarray}
\left ( \begin{array}{cc} \xi_{k_y} & \Delta_{k_y} \\
                          \Delta_{-k_y}^* & -\xi_{k_y}
        \end{array} \right)
\left ( u_{k_y\lambda} \atop{v_{k_y\lambda}} \right ) =
E_{k_y\lambda} \left ( u_{k_y\lambda} \atop{v_{k_y\lambda}} \right)
\end{eqnarray}
where $E_{k_y\lambda}$ with $\lambda = 1 \ldots N$
are the eigenvalues of the Bogoliubov-de Gennes
equations and its eigenvectors $u_{k_y\lambda}(n)$ and
$v_{k_y\lambda}(n)$ are the
coefficients of the Bogoliubov transformation:
\begin{eqnarray}
c_{k_yn\uparrow}&=&\sum_\lambda \left \{
                 \gamma_{k_y \lambda \uparrow} u_{k_y \lambda} (n)
           - \gamma_{k_y \lambda \downarrow}^\dag v^*_{k_y \lambda} (n)
        \right \}
\nonumber \\
c_{-k_yn\downarrow}^\dag&=&\sum_\lambda \left \{
                 \gamma_{k_y \lambda \uparrow} v_{k_y \lambda} (n)
           + \gamma_{k_y \lambda \downarrow}^\dag u^*_{k_y \lambda} (n)
        \right \}
\end{eqnarray}
Furthermore, $\xi_{k_y}$ and $\Delta_{k_y}$ are matrices in the $N$
layers of the $(110)$-wire, i.e. the layers in $x$-direction, and are
given by (for notation see Fig.~\ref{fig:strip110}):
\begin{eqnarray}
(\xi_{k_y})_{nn'} &=& -2t \cos k_y \left(
\delta_{n'n+1}+\delta_{n'n-1}
                                          \right)  -\mu \delta_{n'n}
\nonumber \\
\left (\Delta_{k_y} \right)_{nn'} &=&
      \left( \Delta_{nn'}^+e^{-ik_y}+\Delta_{nn'}^-e^{ik_y} \right)
                                                         \delta_{n'n+1}
\nonumber \\
&+&   \left( \Delta_{nn'}^-e^{-ik_y}+\Delta_{nn'}^+e^{ik_y}
\right)
                                                         \delta_{n'n-1}
\end{eqnarray}
The gap values are determined by the following self-consistency
equations:
\begin{eqnarray}
\Delta_{nn+1}^\pm &=& - V \frac{1}{N} \sum_{k_y} e^{\pm ik_y}
          \langle c_{-k_y n+1 \downarrow} c_{k_y n \uparrow} \rangle
\\
               &=& V \frac{1}{N} \sum_{k_y} \sum_\lambda e^{\pm ik_y}
          u_{k_y\lambda}(n+1) v_{k_y\lambda}^*(n)
\nonumber \label{eq:GapSC}
\end{eqnarray}

To simplify the numerical evaluation of the Bogoliubov-de Gennes
equations we consider isolated wires here. For a Hamiltonian on a
discrete lattice like~(\ref{ham}) an isolated wire is equivalent
to a wire limited by lines of unitary impurities, which is the
boundary condition used for the analytical calculations in the
previous sections.

\begin{figure}[t]
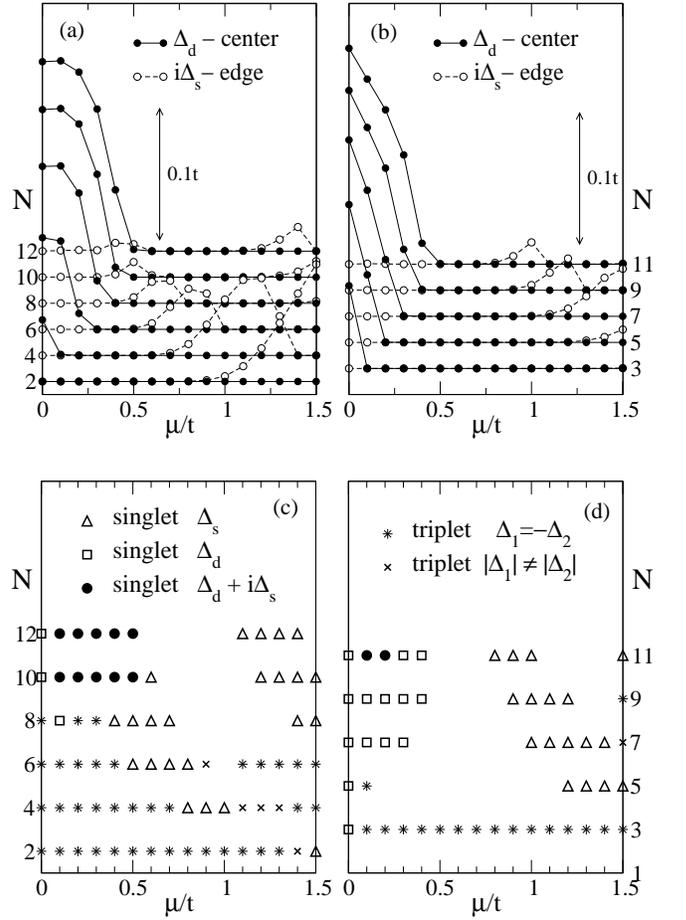

\begin{center}
\includegraphics*[width=.99\columnwidth]{DdsmaxD01.eps}
\\[.5cm]
\includegraphics*[width=.99\columnwidth]{PhaseDiagD01.eps}
\end{center}
\caption{Narrow wires with a nearest neighbor interaction strength of
$V$=1.1575$t$, which gives rise to a gap value of $\Delta_0$=0.2$t$ for
the bulk system at $\mu$=0.
Lower panels: phase diagram of the wires with even width
(c) and with odd width (d). White space
denotes $\Delta=0$, i.e., the normal state.
Upper panels: corresponding amplitudes of the $d_{x^2-y^2}$ and $is$-order
parameters displaying the value in the center of the wire for
the $d_{x^2-y^2}$--wave case and the value at the edge of the
wire in the $s$--wave case. These are the positions where the largest
values of the respective order parameters are expected in the
usual $d_{x^2-y^2}+is$-state.}
\label{fig:PhaseDiagD01}
\end{figure}

\begin{figure}[t]
\begin{center}
\begin{minipage}{.49\columnwidth}
\includegraphics*[width=.98\columnwidth,clip=true]{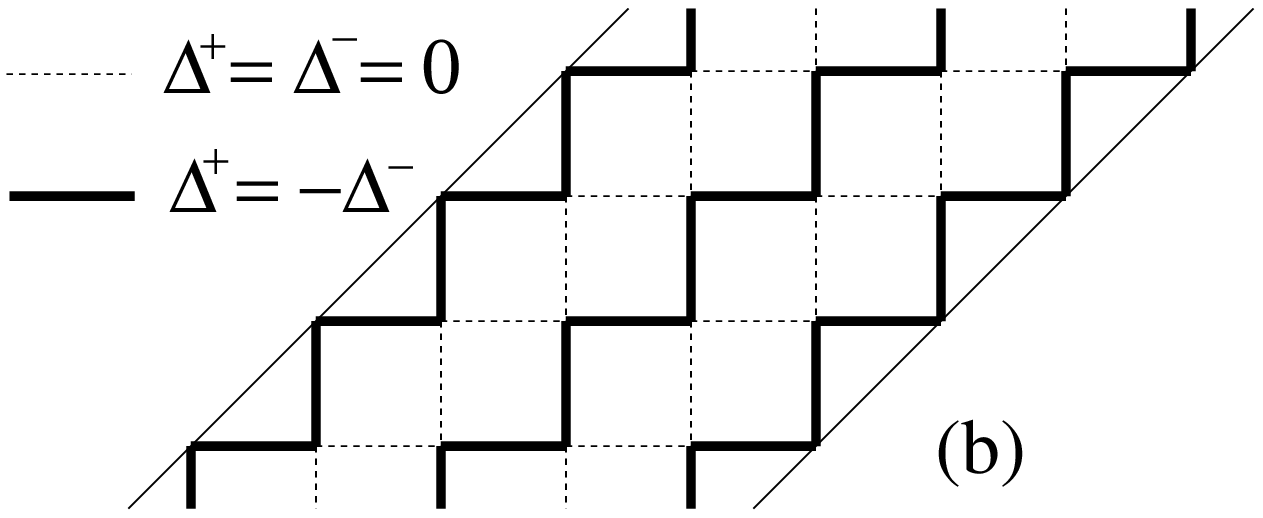}
\\[.3cm]
\includegraphics*[width=.98\columnwidth,clip=true]{DensN08p2.eps}
\end{minipage}
\begin{minipage}{.49\columnwidth}
\includegraphics*[width=.98\columnwidth,clip=true]{DensN08p1.eps}
\end{minipage}
\end{center}
\caption{Quasi one-dimensional triplet superconducting state for $N$=8 and
$\mu$=0. (a) Bond gap values $\Delta^\pm_{nn+1}$ (for definition see
Eq.~(\ref{eq:GapSC})) across the wire.
(b) Illustration of the one-dimensional structure of the
superconducting correlations for an $N$=6-(110) wire at $\mu=0$.
(c) Density of states starting from the outermost
layer of the wire up to the middle layer of the wire, where for
illustrational purposes each
layer has been shifted by an additional offset of 0.3$t$.}
\label{fig:DensN08}
\end{figure}

For narrow wires we find a variety of different phases.
The resulting phase diagram for a nearest neighbor
interaction strength of $V$=1.1575$t$, which gives rise to a gap
of $\Delta_0$=0.2$t$ in a bulk system at $\mu$=0, is displayed in
Fig.~\ref{fig:PhaseDiagD01}. For wires with widths up to $N$=9 we
find a new phase over a wide range of chemical potentials
characterized by $\Delta_{ij}=-\Delta_{ji}$ (stars in
Fig.~\ref{fig:PhaseDiagD01}(c), (d)), which is a signature of
triplet-pairing with $S_z=0$, the only triplet component
compatible with Hamiltonian~(\ref{ham}). For $\mu=0$ the amplitude of
$\Delta_{ij}$ oscillates across the wire between zero and its
maximum value (see Fig.~\ref{fig:DensN08}(a)), indicating a
one-dimensional nature of these new triplet-superconducting
correlations (see Fig.~\ref{fig:DensN08}(b)).
Although the oscillating behavior of $\Delta_{ij}$
remains for finite $\mu$, its amplitude no longer vanishes exactly on
alternating layers. Thus the strict one-dimensionality of the superconducting
correlations seems to be a feature peculiar to $\mu=0$.
This new triplet superconducting phase
will be discussed in more detail below.
At first we will focus on the singlet superconducting phase with possible
$d_{x^2-y^2}$-- and $s$--wave order parameters.

Although we do not understand all details of the variability of
the phase diagram, a few general trends seem clear.
The larger the
width of the wire the more of the usual
$d_{x^2-y^2}+is$--phase (filled circles in
Fig.~\ref{fig:PhaseDiagD01}(c),(d)) is
recovered. The corresponding amplitudes of
the $d_{x^2-y^2}$-- and $is$--components of the order
parameter are displayed in the upper panels of
Fig.~\ref{fig:PhaseDiagD01}. For narrow wires the amplitude of the
$d_{x^2-y^2}$--wave order parameter is finite only for small chemical
potentials $\mu$ and a pure $s$--wave phase
is favorable for large $\mu$. These two phases are separated by a
normal state region.
For larger wire widths the range of the $d_{x^2-y^2}$--wave phase
and the amplitude of the $d_{x^2-y^2}$--wave order parameter
increase. The amplitude of the $s$--wave order parameter, on the other
hand, decreases and the $s$--wave phase moves
towards smaller chemical potentials until it merges with the
$d_{x^2-y^2}$--wave phase, thereby giving rise to a finite
$is$--admixture near the edges of the wire, as expected in
analogy to a single (110)-surface.

\begin{figure}[t]
\begin{center}
\includegraphics*[width=.99\columnwidth]{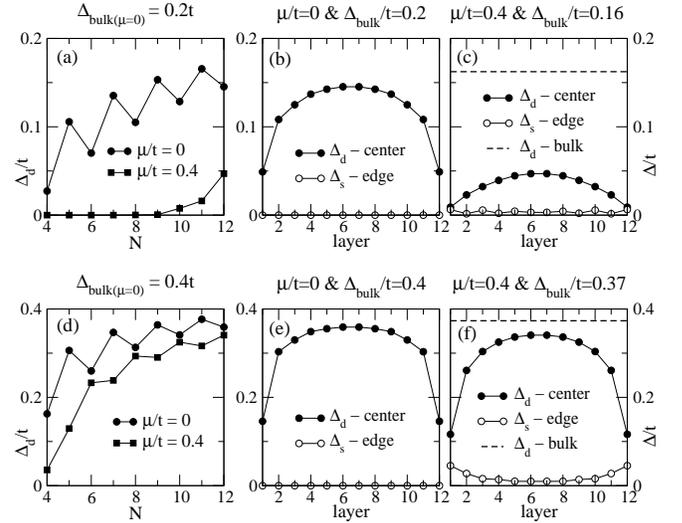}
\end{center}
\caption{Evolution of the $d_{x^2-y^2}$-wave order parameter with
increasing wire width for two different interaction strengths
$V$=1.1575$t$ (upper panels) and $V$=1.7682$t$ (lower panels), which
give rise to a gap value of $\Delta_0$=0.2$t$ and $\Delta_0$=0.4$t$  for the
bulk system at $\mu=0$, respectively. The first column displays
the amplitude of the $d_{x^2-y^2}$-order parameter as a function of the
wire width for two different chemical potentials $\mu/t$=0 and $\mu/t=0.4$.
The other two columns show the variation of the $d_{x^2-y^2}$- and
$is$-component of the order parameter across the wire for $\mu/t$=0
(center column) and $\mu/t$=0.4 (right column). For comparison the
reduced magnitude of the bulk order parameter at $\mu/t$=0.4
is displayed with dashed lines in the right panels.}
\label{fig:D01D02}
\end{figure}

The dependence of the amplitude of the $d_{x^2-y^2}$--wave order
parameter on the width $N$ of the wire is shown in
Fig.~\ref{fig:D01D02}. The upper panels refer to an interaction
strength  of $V$=1.1575$t$, which gives
rise to a gap of $\Delta_0$=0.2$t$ for the bulk system at $\mu$=0.
For $\mu$=0 we observe an even-odd
oscillation in the amplitude of the order parameter (see
Fig.~\ref{fig:D01D02}(a)), which disappears for
larger chemical potentials. The effect of finite $\mu$ on the suppression
of the $d_{x^2-y^2}$--wave order parameter is
considerably stronger in narrow wires than in the bulk system.
For $\mu/t$=0.4 the $d_{x^2-y^2}$--wave order parameter vanishes for
$N<10$ (squares in Fig.~\ref{fig:D01D02}(a)), although the bulk gap
is only reduced by a factor of approx. 0.8 (see
Fig.~\ref{fig:D01D02}(c)).
The dramatic suppression of the $d_{x^2-y^2}$--wave order parameter
for finite $\mu$ is reduced upon consideration of a larger nearest neighbor interaction
strengths, as can be seen in the lower panels of Fig.~\ref{fig:D01D02}.
For a larger interaction strength of $V$=1.7682$t$, which gives rise to
a bulk gap value of
$\Delta_0$=0.4$t$ at $\mu=0$, a pronounced even-odd effect remains at
$\mu/t$=0.4 (see Fig.~\ref{fig:D01D02}(d)) and the
amplitude in the center of an $N$=12-wire is already very close to the bulk
value (see Fig.~\ref{fig:D01D02}(f)) contrary to the smaller
interaction strength, where the amplitude in the center of an
$N$=12-wire is still suppressed by a factor of more than three with
respect to the bulk value (see Fig.~\ref{fig:D01D02}(c)).

\begin{figure}[h]
\begin{center}
\includegraphics*[width=.99\columnwidth]{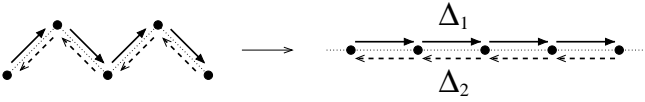}
\end{center}
\caption{Mapping of the $N$=2 wire with $(110)$ orientation to
the 1D-chain with lattice constant $d$. Note that
$\Delta_2=-\Delta_1$ corresponds to a triplet pairing state with
$S_z$=0.} \label{fig:wire1d}
\end{figure}

The emergence of a new triplet-superconducting phase for very
narrow wires can be most easily understood by considering  the
smallest wire, i.e., the $N$=2 wire. Although the effect of
fluctuations will be larger for smaller wires, we focus only on
possible solutions of the BCS mean-field Hamiltonian~(\ref{ham})
in the present paper. Inspection of the phase diagram of
Fig.~\ref{fig:PhaseDiagD01} shows that the $d$-wave order
parameter vanishes for the $N$=2-wire. To investigate alternative
ways in which the $N$=2 wire could lower its ground state energy,
we map it to a 1D-chain (see Fig.~\ref{fig:wire1d}). Although the
gap values $\Delta_{ij}$ and $\Delta_{ji}$ could in principle
differ in amplitude and by a phase factor $\phi$, we restrict our
treatment to $\phi=0$ and $\phi=\pi$. Note that $\phi=0$
corresponds to a singlet pairing state, whereas $\phi=\pi$ is a
triplet pairing state with $S_z=0$\cite{tripletfootnote}. To
investigate the possibility of triplet superconductivity in these
systems in more generality, one should retain pairing correlations
with $S_z=\pm 1$ in the mean-field Hamiltonian as well. For now,
however, we are satisfied with the observation that even for the
simple nearest-neighbor pairing Hamiltonian~(\ref{ham}) a triplet
order parameter can be favored over a singlet order parameter in
narrow geometries.

\begin{figure}[t!]
\begin{center}
\includegraphics*[width=.85\columnwidth,clip=true]{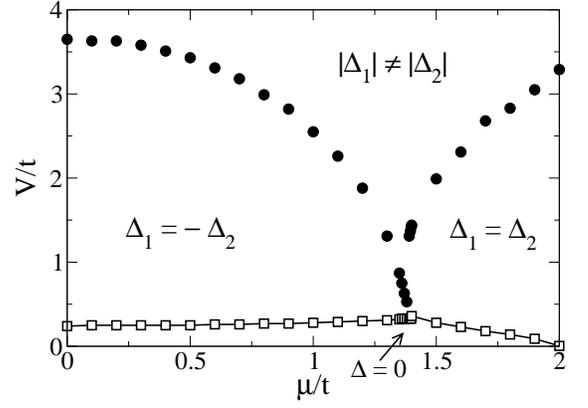}
\end{center}
\caption{Phase diagram for 1D-chain as a function of chemical
potential $\mu/t$ and nearest neighbor interaction $V/t$.}
\label{fig:PhaseDiag1d}
\end{figure}

After Fourier transformation Hamiltonian~(\ref{ham}) for the
1D-chain reads
\begin{eqnarray}
H_{\rm MF}&=&-\sum_{k \sigma} (2t \cos kd + \mu)
                            c_{k \sigma}^\dag c_{k \sigma}
\\
       & &    +\sum_k \left \{ (\Delta_1 e^{ikd}+\Delta_2 e^{-ikd})
                            c^\dagger_{-k \downarrow} c^\dagger_{k \uparrow}
                            + {\rm h.c.} \right\}\,, \nonumber
\label{eq:Ham1d}
\end{eqnarray}
where $\Delta_{ii+1}=-V\langle c_{i+1\downarrow}
c_{i\uparrow}\rangle=\Delta_1$ and $\Delta_{i+1i}=-V\langle
c_{i\downarrow} c_{i+1\uparrow}\rangle=\Delta_2$. It can be easily
diagonalized using the
Bogoliubov-transformation
to give a quasiparticle dispersion of:
\begin{eqnarray}
&E_k&=\sqrt{\epsilon_k^2+ \Delta_k^2} \\
&{\rm with}&
\epsilon_k = \frac{q(k)}{2}+\mu=2t \cos kd + \mu
\nonumber\\
& &\!\!\Delta_k^2 =
\Delta_1^2+\Delta_2^2 +2 \Delta_1 \Delta_2 \cos2kd \,, \nonumber
\end{eqnarray}
where the gap values are to be determined from the following
self-consistency  equations:
\begin{eqnarray}
\Delta_1 &=& - V \frac{1}{4\pi d} \int_{-\pi d}^{\pi d} dk
                    \frac{\Delta_1 + \Delta_2 \cos 2kd}{E_k}
\nonumber \\
\Delta_2 &=& - V \frac{1}{4\pi d} \int_{-\pi d}^{\pi d} dk
                    \frac{\Delta_1 \cos 2kd + \Delta_2}{E_k} \,.
\end{eqnarray}
Four different phases emerge from this model by variation of the
chemical potential $\mu$ and the nearest neighbor interaction
strength $V$ (see Fig.~\ref{fig:PhaseDiag1d}). Below a critical
interaction strength the order parameter vanishes and
the normal state is the ground state of the 1D chain. For
intermediate interaction strengths and small chemical potentials
we find $\Delta_1=-\Delta_2$, i.e., a pure triplet superconducting
state, whereas for large chemical potentials the ground state is
characterized by $\Delta_1=\Delta_2$, i.e., a singlet extended
$s$-wave state. In the singlet state the gap function is $\Delta_k
\sim \cos kd$, i.e., it has nodes at $kd=\pm \pi/2$ whereas the
triplet state gap function is $\Delta_k \sim \sin kd$ and therefore
is maximum at $kd=\pm \pi/2$. This explains why the triplet state
is favored over the singlet state for $\mu=0$, where the Fermi-surface
of the nearest neighbor tight-binding model is at $kd=\pm \pi/2$.
For larger chemical potentials the situation is reversed and the
singlet extended $s$-wave state becomes more favorable than the
triplet state.
For large interaction strengths, we find different magnitudes for
$\Delta_1$ and $\Delta_2$, which corresponds to a mixture of
singlet and triplet pairing. In the limit $V \to \infty$, either
of the gap values $\Delta_1$ and $\Delta_2$ approaches zero
whereas the other goes to infinity, corresponding to an admixture
of triplet and singlet order parameters with equal amplitudes.

With the phase diagram of the 1D-chain in mind we are now able to
better understand the phase diagram of narrow wires as displayed
in Fig.~\ref{fig:PhaseDiagD01}. For small $\mu$ the new phase with
$\Delta_{ij}=-\Delta_{ji}$ simply arises from the formation of
quasi one-dimensional triplet-superconducting correlations along $N$=2
wires (see Fig.~\ref{fig:DensN08}(a),(b)).
This also explains why this new phase is more favorable
for wires with even width than with odd width (compare left and right panels in
Fig.~\ref{fig:PhaseDiagD01}), as only the former can be
divided evenly into $N$=2 wires.
Note that also for the narrow wires,
a critical coupling strength of similar magnitude as in the
1D-chain is necessary to induce the triplet superconducting state,
which is, however, smaller than the
interaction strength considered in this paper.
At larger chemical potentials we find an extended $s$-wave order
parameter analogously to the one-dimensional chain. More difficult
to reconcile with the phase diagram of the one-dimensional chain,
however, are  the seemingly arbitrarily
distributed mixtures of singlet- and triplet-superconducting order
parameters (crosses in Fig.~\ref{fig:PhaseDiagD01}).
The normal state regions, which occur for narrow wires and
intermediate chemical potentials in Fig.~\ref{fig:PhaseDiagD01},
reflect the fact that superconductivity becomes less favorable in
finite geometries and a critical coupling strength is necessary to
induce it \cite{nagai}.

\subsection{Density of states}

\begin{figure}[t]
\begin{center}
\includegraphics*[width=.99\columnwidth,clip=true]{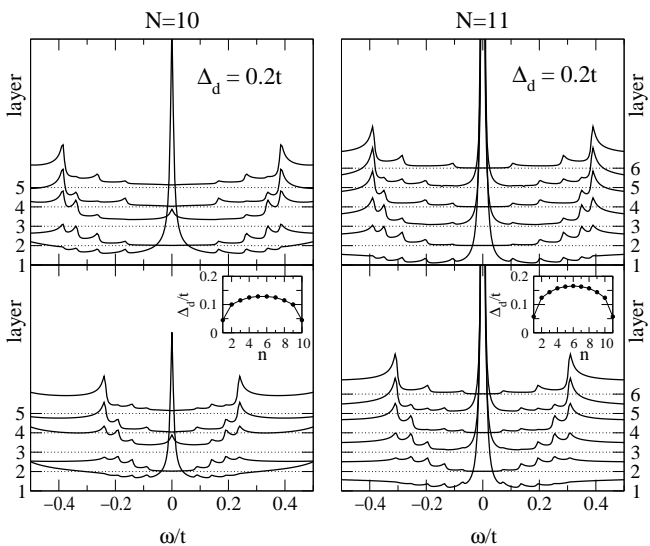}
\end{center}
\caption{Density of states for the $N$=10-wire (left panels) and
the $N$=11-wire (right panels). Upper panels show the
non-selfconsistent results assuming a constant $d_{x^2-y^2}$-order
parameter of $\Delta_0$=0.2$t$, whereas the lower panels display
the density of states resulting from the self-consistently
determined values of the $d$-wave order parameter, whose variation
throughout the wire is depicted in the insets. All panels show the
density of states starting from the outermost layer of the wire up
to the middle layer of the wire, where for each layer towards the
center of the wire an additional offset of 0.3$t$ has been used.}
\label{fig:DensN10N11}
\end{figure}

What does the phase diagram in Fig.~\ref{fig:PhaseDiagD01} imply
about the possible existence of Andreev bound states in narrow
wires? In Fig.~\ref{fig:DensN08}(c) the local density of states and
the variation of the order parameter throughout the wire is
depicted for the quasi one-dimensional triplet superconducting
state using the $N$=8-wire at $\mu$=0 as an example. Obviously,
the density of states in this state is fully gapped, in analogy to
the one-dimensional chain, and there are no Andreev bound states.

Close to half-filling, for wider wire widths, the results are more
conventional and the effects of self-consistency much simpler.  In
Fig.~\ref{fig:DensN10N11} the local density of states is displayed
for the $N$=10- and the $N$=11-wire at $\mu=0$, which are
characterized by a pure $d_{x^2-y^2}$-order parameter. Here, the
main difference between the self-consistent (lower panels of
Fig.~\ref{fig:DensN10N11}) and the non-self consistent results
(upper panels of Fig.~\ref{fig:DensN10N11}) is the suppression of
the magnitude of the $d$-wave order parameter especially towards
the edges of the wire (see also insets of
Fig.~\ref{fig:DensN10N11}). Whereas the  weight of the zero-energy
peak is reduced in the even-width ($N$=10) wire the weight of
the zero-energy state in the odd-width wire ($N$=11)
is hardly affected.

\section{conclusion}

\label{sec:conclusion}

Motivated by recent scanning tunneling experiments on the
Ba$_2$Sr$_2$CaCu$_2$O$_8$ systems which reveal inhomogeneous
electronic structure on the nanoscale, we have analyzed in detail
the electronic structure of $d$-wave quantum wires. These wires
exemplify the effects of a constrained geometry while still being
simple enough to allow for analytical solutions. To impose a
restricted geometry we use lines of impurities with infinite
scattering strength, a method which allows to cut arbitrarily
shaped objects out of the two-dimensional plane. In principle, it
is straightforward to extend this method to investigate the
effects of tunneling between neighboring grains by reducing the
scattering strength of the impurities and thus lowering the
potential barrier between neighboring grains.

New and interesting physics arises from interference effects
between the two surfaces of the wire when its width is of the
order of the superconducting coherence length. Contrary to
$s$-wave superconductors in finite geometries, the surface pair
breaking plays an important role. In this respect, the existence
and nature of Andreev bound states in constricted geometry is of
particular interest. In order to single out new effects peculiar
to quantum wires and arising from the interference of the two
surfaces, we have first addressed the case of a single surface in
the first part of the paper, concentrating on surfaces with
$(100)$-, $(210)$- and $(110)$-orientations in a nearest neighbor
tight-binding model at half-filling. Andreev bound states can form
on surfaces with orientations deviating from the $(100)$ direction
due to the sign change of the $d$-wave order parameter. However, in
the presence of several channels for reflection of quasiparticles
from the surface, the zero-energy Andreev states may not exist, as
 is the case for the $(210)$ surface of the square lattice. Here
our results are in qualitative agreement with earlier work based
either on the quasiclassical approximation or Bogoliubov-de Gennes
equations. Only for the $(110)$-surface, which involves the
strongest pair-breaking, do we find a zero-energy Andreev bound
state, whose amplitude decreases with the square of the inverse
distance from the surface and vanishes on even layers.

In the main part of this work, we  focussed on quantum wires with
$(110)$-orientation, which
display a pronounced ``width parity" effect.  The special case of
electrons hopping on a square lattice with half-filled
tight-binding band was treated most extensively.  For wires of
this type with an odd number of chains $N$ only one subgap state,
a dispersionless zero-energy Andreev-bound state, exists. As in
the single surface case, the amplitude of the ZES vanishes on even
layers and decays towards the center of the wire. In addition to
the zero-energy state (ZES), there are $N$-1 dispersive modes,
which are doubly degenerate due to particle-hole symmetry and are
not Andreev states.

In the case of even $N$ half-filled wires, a splitting of the
branches occurs which is associated with a symmetry breaking $k_y
\to - k_y$, and a total of 2$N$ dispersive modes exist. Although
there is no dispersionless ZES,  all quasiparticle states in the
even width wire are of Andreev character in the sense that the
current of the probability density vanishes due to opposite
contributions from particle and hole excitations.  These
quasiparticles occupy either conventional Andreev-type surface
states, or a new type of Andreev standing wave, according to their
momentum $k_y$ parallel to the wire. In the 2D limit $N \to
\infty$,  the lowest energy dispersive state was shown to
transform into the usual zero-energy Andreev bound state at the
impenetrable surface, while the standing wave states evolve
into either the continuous spectrum or the surface states. With
increasing deviation from  half-filling, odd-even effects in
the wires become less pronounced. The evolution of the Fermi surface
shape with these deviations can result in additional extrema in
the quasiparticle dispersive modes and, hence, new peaks in the
LDOS. Large-scale faceting of the surfaces with characteristic
scales larger than the wire thickness will not influence our
results significantly. However, small-scale inhomogeneities like point-like
defects and impurities can substantially change the effects of
interference induced by wire surfaces, even if the phase breaking
length is large.

The small coherence length of high-temperature superconductors
leads to further restrictions on the applicability of the
quasiclassical results to superconducting wires or films, whose width is
less than or comparable to the coherence length. For narrow wires of
$(110)$-orientation, the self-consistent treatment of the
order-parameter is found to have a large effect. For wires with
widths less than the superconducting coherence length (up to $N$=9
for the particular parameters used in our numerical calculations),
especially for the even width wires, a new phase characterized by
quasi one-dimensional triplet pairing is found in the mean-field
phase diagram. This new phase is fully gapped and characterized by
the absence of Andreev bound states. The larger the width of the
strip, the more the $d+is$-state, which is expected for a single
surface with $(110)$-orientation, dominates the phase diagram near
half filling. With respect to the density of states at
half-filling, where the order parameter has only a $d$-wave
component, the main effect of the self-consistent treatment is the
suppression of the magnitude of the $d$-wave order parameter,
especially near the surfaces of the wire.

It is interesting to end this discussion with some speculations on
the role of bound quasiparticle states and edge effects of this
type on the spectra of weakly coupled superconducting grains as
apparently observed in STM experiments. Such irregular grains
should contain nanoscale ``facets" at all possible angles, so
presumably the most general situation with sizeable particle-hole
asymmetry and mixture of even- and odd-$N$ boundary conditions
will apply. If we first assume that the pair interaction and grain
size are such that one may ignore the triplet states found in the
self-consistent treatment, we expect the spectra of weakly coupled
grains to be dominated by the zero-dimensional analogs of the
dispersive subgap states discussed here, i.e. there should be a
wide distribution of bound state energies depending on local
geometry of the grain, and visible in the LDOS as measured by STM.
In this sense we question whether the weakly-coupled grain picture
is in fact applicable to the experiments in question, which appear
to see a very {\it homogeneous} spectrum  at low energies in the
superconducting state.

A second remark is based on our observation that in nanoscale
confined geometry, spin triplet fluctuations may become more
favorable. Such time-reversal symmetry breaking fluctuations will
clearly lead to local spontaneous currents, an issue which has
recently been raised again in angle-resolved photoemission
studies\cite{CampuzanoTbreaking}. Future studies of small grains
are planned to address these issues.

\section{Acknowledgments}
We thank I.~Bobkova, Ya.~Fominov, T.~Kopp, J.~Mannhart and K.~Ziegler for useful
discussions. This work was supported, in part, by grant RFBR 02-02-16643
(A.M.B. and Yu.S.B.) and NSF-INT-0340536 (P.J.H. and Yu.S.B). A.M.B.
acknowledges the support of Dynasty Foundation and T.S.N.
the support of the Alexander von Humboldt Foundation.


\begin{thebibliography}{stuffstuffstuff}
\bibitem{ebz} In the presence of half-filling, surfaces with $(110)$
orientations represent a special case for the square lattice, since
the parallel component of the Fermi momentum coincides in this case with
the edge of the surface-adapted Brillouin zone. In the problems we study,
the role of the edge of the Brillouin zone becomes quite important, in
particular, for this reason. In a more general case one can expect, that
extrema of the quasiparticle spectra would arise at other values of the
momentum component parallel to the surface, which are comparable with
the scale of the Brillouin zone.
\bibitem{tripletfootnote} In terms of standard representation of singlet and triplet pairings
$\hat{\Delta}_s=\Delta i\hat{\sigma}_2$, $\hat{\Delta}_t=(\bm{d}\cdot\hat{\bm{\sigma}})i\hat{\sigma}_2$
one finds for the order parameters we introduced $\Delta_1=\Delta+d_z$, $\Delta_2=\Delta-d_z$,\, $d_x=d_y=0$.
Hence, in the purely singlet case $\Delta_1=\Delta_2$, while for the triplet pairing $\Delta_1=-\Delta_2$.
Otherwise there is a mixture of singlet and triplet pairings.
\end{thebibliography}
\end{document}